\documentclass[aps,pre,twocolumn,showpacs]{revtex4}
\usepackage{amsmath,graphicx,amsbsy,amssymb,epsfig,latexsym,wasysym,pifont,mathrsfs} 
\usepackage{float}
\usepackage{color}
\usepackage{xcolor}
\usepackage{hyperref}
\usepackage{multirow}
\usepackage{comment}
\begin{document}

\title{Pattern formation and phase transition in the collective
  dynamics of a binary mixture of polar self-propelled particles}

\author{Sagarika Adhikary}%
\author{S. B. Santra}%
\affiliation{Department of Physics, Indian Institute of Technology
Guwahati, Guwahati-781039, Assam, India.}


\begin{abstract}
The collective behavior of a binary mixture of polar self-propelled
particles (SPPs) with different motile properties is studied. The
binary mixture consists of slow-moving SPPs (sSPPs) of fixed velocity
$v_s$ and fast-moving SPPs (fSPPs) of fixed velocity $v_f$. These SPPs
interact via a short-range interaction irrespective of their types.
They move following certain position and velocity update rules similar
to the Vicsek model (VM) under the influence of an external noise $\eta$.
The system is studied at different values of $v_f$ keeping $v_s=0.01$ constant
for a fixed density $\rho=0.5$. Different phase-separated collective patterns
that appear in the system over a wide range of noise $\eta$ are characterized.
The fSPPs and the
sSPPs are found to be orientationally phase-synchronized at the steady-state.
We studied an orientational order-disorder transition varying the angular
noise $\eta$ and identified the critical noise $\eta_c$ for different $v_f$.
Interestingly, both the species exhibit continuous transition
for $v_f<100v_s$, and discontinuous transition for $v_f>100v_s$.
A new set of critical exponents is determined for the continuous transitions.
However, the binary model is found to be non-universal as
the values of the critical exponents depend on the velocity. The
effect of interaction radius on the system behavior is also studied.
\end{abstract}

\pacs{05.70.Fh,87.18.Gh}

\maketitle
 
\section{Introduction}
Collective pattern formation and self-organization of active or
self-propelled particles (SPPs) spontaneously occur in nature at
different length scales. For example, actin filaments can form
different structures as micro-clusters, stripes, and traveling waves
\cite{schaller2010polar, kohler2011structure}, bacterial swarms
growing in the laboratory can self-organize into high-density coherent
patterns \cite{murray2001mathematical,
  thar2005complex,sokolov2007concentration,
  zhang2009swarming,zhang2010collective}; other examples include
army ants \cite{couzin2003self}, insect swarms
\cite{buhl2006disorder,romanczuk2009collective}, spiny lobsters
\cite{herrnkind1969queuing, bill1976drag}, fish schools
\cite{makris2006fish}, flocking birds \cite{ballerini2008interaction,
  cavagna2010scale}, pedestrian flow
\cite{helbing1995social,jiang2017extended}, artificial systems of SPPs
\cite{narayan2007long,deseigne2010collective}. The collective behavior
of polar SPPs is extensively studied following the seminal
Vicsek Model (VM) \cite{vicsek1995novel}. In this model, a large
number of polar SPPs move together with equal speed $v_0$, and they
align their direction of motion with their neighbors through a
short-range ($R$) alignment interaction. However, the average
direction is subject to an angular noise ($\eta$). For a given density
($\rho_0$), an orientational order-disorder transition occurs at a
critical noise ($\eta_c$). Initially, the nature of this phase
transition in the VM was found to be continuous for low velocity on
small system sizes\cite{nagy2007new, vicsek2012collective}. However,
later it is established through extensive simulations that there
exists a crossover system size $L^*(\rho_0,v_0)$
\cite{chate2008,adhikary2021effect} below which the nature of the
transition is continuous, and above which it is discontinuous where
dense traveling bands appear in the system. It needs to be noted that
$L^*(\rho_0,v_0)$ diverges both for low velocities $(v_0<0.05)$ and
low densities $(\rho_0<0.01)$ \cite{chate2008}. The formation of the dense
traveling band near the transition region is fluctuation-driven
and occurs due to the feedback mechanism between local order and local
density \cite{ginelli2016}.

One of the major limitations of the VM is that all the polar SPPs have
the same motile property. For example, all the SPPs have the
same velocity. However, in natural systems, the velocities of
particles need not be the same during collective motion. For example,
there are fast-moving (active) and slow-moving (dormant) bacteria in a
bacterial population or slow-moving and speedy vehicles in daily
traffic. Phase-separation and static clusters were mainly observed in the
study of collective dynamics of SPPs with variable
speed (depending on the neighborhood's polarization)
\cite{mishra2012collective,singh2020phase}. Apart from velocity, other
properties of active particles are also varied to study the phase
separation in different systems
\cite{cates2015motility,bechinger2016active}. Examples of such
systems include: a mixture of active Brownian
particles with different diffusion constant
\cite{weber2016binary,demix_sunita}, a mixture of active and passive
particles \cite{stenhammar2015activity, dolai2018phase,
  maloney2020clustering, mccandlish2012spontaneous}, binary active
particles with different alignment interactions \cite{menzel},
an oppositely driven binary mixture of particles
\cite{reichhardt2018laning, Ikeda_2012, bain2017critical}, chiral
active matter \cite{article, liebchen2017collective}, a mixture of
polar and apolar SPPs \cite{sampat2021polar} and many others.

However, the study of collective dynamics in a binary mixture of polar
SPPs with different motile properties is a new area of research. This
paper studies the collective dynamics of a mixture of
two types of SPPs with widely different fixed velocities. The SPPs
interact with local short-range interaction, and the dynamics evolve
under an external noise. It is intriguing to observe whether 
self-organized pattern formation and phase separation occur in this
binary model (BM) with a short-range interaction only. Furthermore,
will there be any criticality in
the system? Will the system undergo any order-disorder transition
at a critical point? What would be the nature of such transition? In
this study, we explore answers to all these questions. The collective
dynamics and the underlying mechanism could also explain the similar
pattern formation and dynamical behavior observed in other systems. In
the next section, we present the model and discuss the results
obtained. Herein, polar SPPs will be
referred to simply as SPPs.

\section{Model}
\begin{figure}[t]
\centerline{
  \hfill\psfig{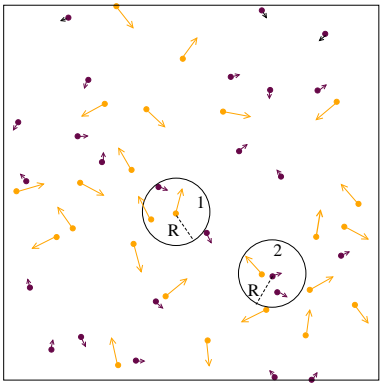}
\hfill }
\caption{\label{model} The distribution of binary SPPs is shown on a
  system of size $L=10$, where $N=\rho L^2=50$, $N_f=N_s=25$. Orange and maroon
  colors represent fSPPs and sSPPs, respectively. The arrow
  associated with an SPP indicates the direction of velocity. A
  fSPP at the center of the circle-$1$ interacts with both the fSPP
  and sSPP present within the radius of $R$. Similarly, an sSPP at
  the center of the circle-$2$ interacts with both types of
  SPPs present within the radius $R$.}
\end{figure}

The collective motion of a mixture of SPPs with two different motilities is
modeled over a two-dimensional square box of linear size $L$. The
different motilities of SPPs are modeled, assigning widely different
fixed velocities to them. The fast-moving SPPs (fSPPs) move with a velocity
$v_f$ and the slow-moving SPPs (sSPPs) move with a velocity $v_s$ where
$v_f\gg v_s$.  The two types of SPPs are taken in equal proportion. If
$N_{f}$ is the number of fSPPs, and $N_{s}$ is the number of
sSPPs, then $N_{f}=N_{s}=N/2$ where $N$ is the total number
of SPPs in the system. Initially, the position $\vec{r}_{p,i}$,
$i=1,2,3,\cdots,N/2$ of all the SPPs are randomly distributed over the
space (off-lattice) where $p\in\{s,f\}$. The initial orientation
$\theta_i$ of an SPP is randomly selected in the range $-\pi$ to
$\pi$, irrespective of their type. The SPPs of both types interact
within a local neighborhood $R=1$ and determine their average
orientation. There is no external force or inter-particle repulsion.
Both inter-particle and intra-particle interactions are
considered in estimating the average orientation. The distribution of
randomly oriented $25$ fSPPs (in orange) and $25$
sSPPs (in maroon) are shown in Fig.\ref{model}. Longer and
shorter arrows show the velocities $v_f$ and $v_s$, respectively.

The time evolution of the orientation $\theta_{i}$ of the
$i$th SPP is determined by
\begin{equation}
  \label{d1}
  \theta_{i}(t+\Delta t) = {\langle\theta(t)\rangle}_{R} + \Delta\theta
\end{equation}
where $\Delta\theta$ is a random orientation chosen with a uniform
probability from the interval $[-\eta\pi,+\eta\pi]$. The strength of
the angular noise $\eta$ varies from $0$ to $1$. The term
$\langle\cdots\rangle_{R}$ is defined as
\begin{eqnarray}
  \label{da}
 & {\langle\theta(t)\rangle}_{R} = \arctan
  \left[\frac{\langle\sin\theta\rangle}{\langle\cos\theta\rangle}\right];
  \nonumber \\ &\langle\sin\theta\rangle= \frac{1}{n} \sum_{j\in\{R\}}
  \sin\theta_j; \ \ \langle\cos\theta\rangle= \frac{1}{n}
  \sum_{j\in\{R\}} \cos\theta_j \nonumber
\end{eqnarray}
where $n$ is the number of SPPs in the neighbourhood region $R$ that
includes both the sSPPs and fSPPs. It should be noted that
the magnitude of velocity of individual SPPs are ignored and only the
orientations are taken into account in estimating
$\langle\theta(t)\rangle_R$. After averaging, an SPP of type-$p$
($p\in\{s,f\}$) at the position $\vec{r}_{p,i}$ is thus moving with a
speed $v_p$ in the direction $\theta_i$. Knowing the
velocity $\vec{v}_{p,i}(t)$ at every time step, the position of the
$i$th SPP $\vec{r}_{p,i}$ is updated following the forward update rule
\begin{equation}
  \label{d2}
{\vec{r}}_{p,i}(t+\Delta t) = {\vec{r}}_{p,i}(t) +
{\vec{v}}_{p,i}(t)\Delta t
\end{equation}
where $\Delta t$ is the time between two successive updates, and it is
chosen as $\Delta t=1$. Periodic boundary conditions (PBCs) in the
horizontal and vertical directions are applied in case the position
vector crosses the boundary of the square box. Eq.\ref{d1} and
Eq.\ref{d2} are then evolved with time and a steady state is achieved
for a given noise $\eta$. The model is studied varying $\eta$ over a
wide range. The interaction considered in this model is purely
aligning in nature and no steric or hydrodynamic interactions present
in the system. In a special situation if $v_s=v_f=v_0$, the BM is
equivalent to the VM with velocity $v_0$ for all SPPs
\cite{vicsek1995novel}.

\begin{figure*}[t]
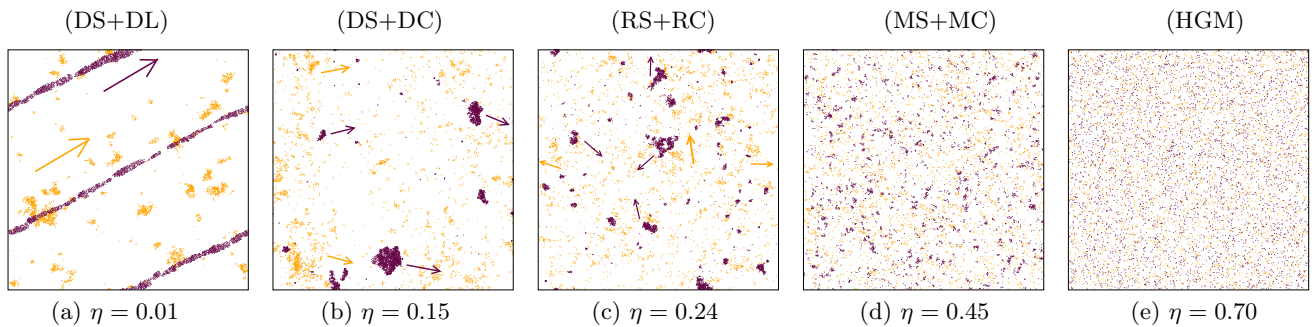

   \centerline{\hfill(DS+DL) \hfill\hfill(DS+DC)
   \hfill\hfill(RS+RC) \hfill\hfill(MS+MC)
   \hfill\hfill(HGM) \hfill}
  \vspace{0.5em}
   \centerline{
      \hfill\psfig{file=fig2a.eps,width=0.18\textwidth}
       \hfill\psfig{file=fig2b.eps,width=0.18\textwidth}
      \hfill\psfig{file=fig2c.eps,width=0.18\textwidth}
      \hfill\psfig{file=fig2d.eps,width=0.18\textwidth}
       \hfill\psfig{file=fig2e.eps,width=0.18\textwidth}
       \hfill }
 \centerline{\hfill(a) $\eta=0.01$ \hfill\hfill(b) $\eta=0.15$
   \hfill\hfill(c) $\eta=0.24$ \hfill\hfill(d) $\eta=0.45$
   \hfill\hfill(e) $\eta=0.70$ \hfill}
\caption{\label{morph_all} System morphology with
  $v_f=30v_s$ and $v_s=0.01$ for different angular noise (a)
  $\eta=0.01$, (b) $\eta=0.15$, (c) $\eta=0.24$, (d) $\eta=0.45$ and
  (e) $\eta=0.70$ for a system of size $L=128$, $N_f=N_s=4096$.
  Orange: fSPPs and maroon: sSPPs. }
\end{figure*}

 For a given initial random distribution of SPPs over a
  square box of size $L$, Monte Carlo simulations are performed to
  study the time evolution of the binary system following Eq.\ref{d1}
  and Eq.\ref{d2}. One Monte Carlo time step corresponds to
  incrementing the position and orientation of all the SPPs. The initial
  $7\times 10^5$ Monte Carlo steps are neglected to achieve the
  steady-state. An ensemble of size $48\times10^5$ is taken for
  statistical averages ($2\times 10^5$ samples at different times for
  $24$ different initial configurations). Simulations are performed
  at different velocities: keeping $v_s=0.01$ fixed, the value of
  $v_f$ is taken as $v_f=30v_s$, $v_f=50v_s$, $v_f=100v_s$ and
  $v_f=150v_s$. The density of SPPs is kept fixed as $\rho=0.5$ for
  all simulations. Thus, the total number of SPPs is $N=\rho L^2$. The
  number of sSPPs ($N_s$) and fSPPs ($N_f$) are taken in equal
  proportions, {\em i.e.}; $N_f=N_s=N/2$.

\section{Collective Patterns and its characterization}
In this section, we will describe the morphological as well as
structural changes of flocks of SPPs with noise. A flock is a
collection of SPPs who are within the range of interaction $R$ from
each other. For this study, we consider the velocity of sSPPs as
$v_s=0.01$ and that of fSPPs as $v_f=30v_s$. The morphology of the
binary mixture at different noises is generated on a system of size
$L=128$ (total number of SPPs $N=8192$,
$N_f=N_s=4096$). The morphology is shown in Fig.\ref{morph_all} where
the orange color represents the fSPPs, and the maroon color represents the
sSPPs. A variety of patterns of the flocks of sSPPs and fSPPs appear
as the angular noise $\eta$ varies.

For a low angular noise $\eta=0.01$, the flocks of fSPPs are in the
form of clusters and moving in a particular direction (indicated by an
orange arrow). On the other hand, the flocks of sSPPs form narrow
lanes in the same direction (indicated by a maroon arrow) as that of
fSPPs. The situation is shown in Fig.\ref{morph_all}(a). It seems that
in the binary system of SPPs, the dynamics of sSPPs is governed by the
local shear between two SPPs. Since the sSPPs are slow movers, many
fSPPs interact with them at a given time. Moreover, at each time
step, the fSPPs travel a much larger distance than the sSPPs, and
consequently, they influence the ordering of the distant flocks of
fSPPs as well as that of sSPPs. For a given initial configuration, the
direction of motion of directed clusters (DS) of fSPPs or directed
lanes (DL) of sSPPs, is spontaneously selected toward an arbitrary
direction. Such lane formation was not observed in the VM with
monodispersed SPPs. However, the formations of lanes and clumps were
also observed in a model of SPPs with
density-dependent motility \cite{farrell2012pattern}. 
Phase-separated bands of big and small particles oriented along the
direction of the flow are also observed in a bidisperse granular system
under shear flow \cite{santra1996fluid}.

As $\eta$ is increased to $0.15$, DSs of fSPPs look a little
scattered, whereas the sSPPs form clumps (or compact clusters) that
still move in the same direction as DSs as shown in
Fig.\ref{morph_all}(b). The orientation of the clumps of sSPPs remains
synchronized with that of the DS of fSPPs. We call these clumps of
sSPPs, directed clumps (DC). Further increase of noise to $\eta=0.24$
induces random motion to the flocks of both fSPPs and sSPPs, shown in
Fig-\ref{morph_all}(c). We call the randomly moving clusters of fSPPs
by random clusters (RS) and the randomly moving clumps of the sSPPs by
random clumps (RC). Thus, the orientationally ordered phase with DS
and DC goes to an orientationally disordered phase with RS and RC as
the noise changes from $\eta=0.15$ to $\eta=0.24$. The critical
behavior of such a transition will be described in Section IV. As $\eta$ is
increased to $0.45$, RS and RC are found to dissolve into a larger
number of micro-clusters (MS) of fSPPs and micro-clumps (MC) of sSPPs,
respectively, as shown in Fig.\ref{morph_all}(d). For relatively high
$\eta$, a homogeneous gas-like mixture (HGM) of both types of SPPs is
observed as shown in Fig.\ref{morph_all}(e) for $\eta=0.70$.

Similar patterns also appear in the cases of higher values of $v_f$
keeping $v_s=0.01$. Now we characterize the patterns considering a
system of size $L=256$ with $\rho=0.5$. Thus, the total
number of SPPs $N=32768$, and the numbers of fSPPs and sSPPs are
$N_f=N_s=16384$.

\subsection{Phase segregation}
\begin{figure}[t]   
   \centerline{\psfig{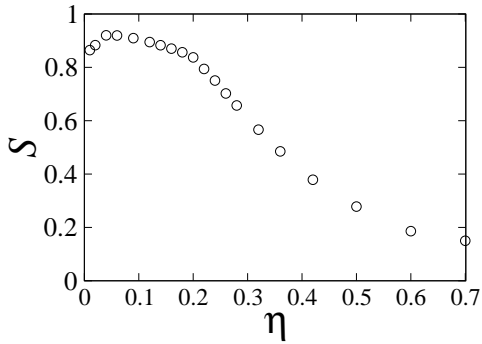}}
   \caption{\label{pattern2} $v_{f}=30 v_s, v_s=0.01$: Plot of the
     segregation coefficient $S$ versus $\eta$.  System size is
     $L=256$. The square box of area $L^2$ is divided into $Q$ square
     sub-regions of linear size $\ell$ and area $L^2/Q$ each.}
\end{figure}

It is observed that the system remains phase-separated over a wide
range of noise $\eta$. The system is found to be phase-separated,
which means the sSPPs and fSPPs can be identified separately in space. 
A segregation coefficient \cite{article,yang2014aggregation} is defined as
\begin{equation}
  \label{d_seg}
  S = \frac{1}{N}\sum_{j=1}^{Q}\left|{n^{(j)}_{f}-n^{(j)}_{s}}\right|
\end{equation}
where $\sum_{j=1}^{Q}\{n^{(j)}_{f}+n^{(j)}_{s}\}=N$, the total number
of SPPs, $Q$ is the number of small square boxes of size
$\ell\times\ell$ that cover the whole system of size $L\times L$,
$n_{f}$ and $n_{s}$ are the number of the fSPPs and sSPPs,
respectively, in a small box. We have estimated $S$ for a wide range
of $\eta$ taking $\ell=8$. Thus, $Q=L^2/\ell^2=1024$.  As per
Eq.\ref{d_seg}, if every box is filled with either fSPPs or sSPPs
(phase-separated), then $S=1$ and if every box is filled with both
fSPPs and sSPPs in equal proportion (homogeneous phase), then
$S=0$. Whereas $S=1/2$ corresponds to the situation that, on an
average, $\Delta n=\left|{n^{(j)}_{f}-n^{(j)}_{s}}\right|\approx 16$
in every box. We consider the system to be phase-separated if
$S\ge1/2$ and mixed if $S<1/2$.  The variation of $S$ with $\eta$ is
shown in Fig-\ref{pattern2}. It can be seen that the system remains
phase-separated up to a very high noise $\eta\approx0.36$.
The value of $S$ is greater than $0.8$ for $\eta<0.04$ at which the
phase-separated DL and DS appear. The process of phase separation
continues, and the system is found to be highly phase separated with
$S\approx0.9$ in the noise range $0.04<\eta<0.09$. Such a phase
separation between the two species is due to the large
velocity difference and the presence of noise in the
system. For $\eta>0.09$, $S$ decreases monotonically with increasing
$\eta$. This indicates a slow mixing of the two species of SPPs at
high noise. Eventually, at $\eta\approx 0.70$, the phase separation
disappears $S\approx0.15$, and the two species of SPPs mix
uniformly. Phase separation was also reported in a binary mixture of
active Brownian particles with wide differences in the diffusion
constant \cite{weber2016binary}.

\subsection{Characterization of patterns}
\begin{figure}[t]
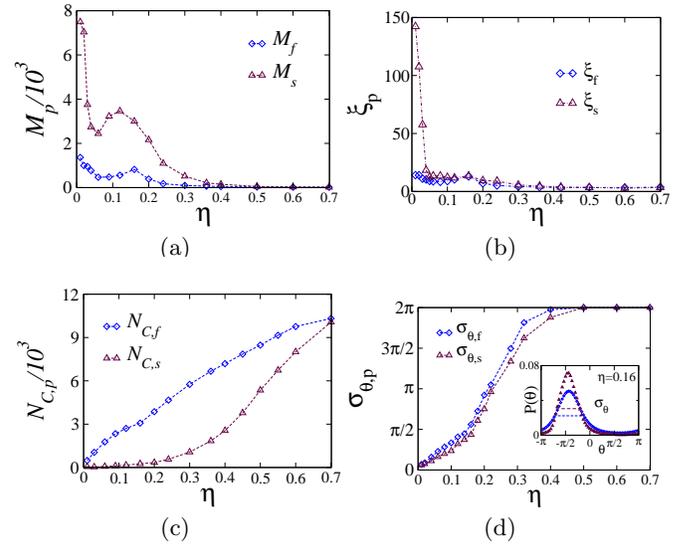
   
  \centerline{
    \hfill\psfig{file=fig4a.eps,width=0.24\textwidth}
        \hfill\psfig{file=fig4b.eps,width=0.24\textwidth}
    \hfill }
  \centerline{\hfill (a) \hfill\hfill (b) \hfill}
     \vspace{1.0 em}
     \centerline{\hfill
   \psfig{file=fig4c.eps,width=0.24\textwidth}
     \hfill\psfig{file=fig4d.eps,width=0.23\textwidth}
    \hfill }
   \centerline{\hfill (c) \hfill\hfill (d) \hfill}
  \caption{\label{pattern1} $v_{f}=30 v_s, v_s=0.01$: (a) Plot of
    $M_p$ against $\eta$. (b) Plot of $\xi_p$ against $\eta$.  (c)
    Plot of $N_{C,p}$ against $\eta$. (d) Plot of $\sigma_{\theta,p}$
    versus $\eta$. In the inset of (d), $P(\theta)$ is plotted for
    $\eta=0.16$ with the same symbols and colors for the fSPPs and
    sSPPs. System size is $L=256$.}
\end{figure}

To monitor the structural change in the patterns of the flocks, we
have estimated the size ($M_P$) and extension ($\xi_p$) of the largest
flock, and the number of flocks $N_{c,p}$, ($p\in\{s,f\}$) with noise
$\eta$ on a system of size $L=256$. The largest flock size of $p$-type
SPPs is given by the number of $p$SPPs that belong to the largest
flock of that type, and the extension of a flock is just the gyration
radius. To measure how the flocks of SPPs are ordered, we have
estimated the full-width at half maxima $\sigma_p$ of the orientation
distribution $P(\theta)$ where $\sigma_p$ can vary from $0$ to
$2\pi$. Thus, $\sigma_p$ represents the orderliness of the flocks,
$\sigma_p=0$ means they are highly ordered (moving in a particular
direction), and $\sigma_p=2\pi$ means they are completely disordered
(moving in all possible directions). The variations of $M_p$, $\xi_p$,
$N_{c,p}$ and $\sigma_p$ against $\eta$ are shown in
Fig.\ref{pattern1}.

For $\eta=0.01$, the sSPPs have the largest flock of size $M_s\approx
8000$, extension $\xi_s\approx 150$ (half of the system size) and
number of flocks $N_{c,s} \approx 50$. Whereas, the fSPPs have the largest
flock of size $M_f\approx 1300$ (out of $N_f=16384$), extension
$\xi_f\approx 15$ and number of flocks $N_{c,f}\approx 500$. At this noise,
the flock structure of sSPPs (DL) seems to be more compact in
comparison to the DS of fSPPs. It should be noted that such a dense
packing of slow-moving particles occurred in the absence of
an attractive force. This has happened only due to the presence of
small noise and high-velocity differences between the two types. The
flocks are highly ordered, $\sigma_s=\sigma_f\approx\pi/15$.  

As $\eta$ is increased $0.06$, a drastic change happened to the flock
structure of sSPPs. The largest flock size of sSPPs is $M_s\approx
2500$, the extension is $\xi_s\approx 12$, and the number of flocks is
$N_{c,s}\approx 70$. For fSPPs, the largest flock size is
$M_f\approx500$, the extension is $\xi_f\approx10$ and number of
flocks is $N_{c,f}\approx 1700$. The change in the structure of sSPPs
is remarkable in comparison to that of fSPPs. It seems there is a
noise-induced melting transition happening in the system. However, it
needs to be characterized carefully with further studies. The
orderliness of the flocks are $\sigma_s=\sigma_f\approx\pi/6$.

On further increase of $\eta$ to $0.12-0.16$, both sSPPs and fSPPs
rebuild their largest flocks. For sSPPs at $\eta=0.12$, the largest
flock size is $M_s\approx3400$, the extension is $\xi_s\approx12$. The
number of flocks has increased to $N_{c,s}\approx160$. The orderliness
$\sigma_s\approx\pi/3$. For fSPPs at $\eta=0.16$, the largest flock
size is $M_f\approx 800$, the extension is $\xi_f\approx12$. The
number of flocks has increased to $N_{c,f}\approx3000$. The
orderliness $\sigma_f\approx\pi/2$. In the intermediate range of
$\eta$, it is observed that $\sigma_f>\sigma_s$. Once again it should
be noted that the aggregation of flocks is happening in absence of any
attractive force. The orientation distribution $P(\theta)$ for both
the SPPs are given in the inset of Fig.\ref{pattern1}(d) for
$\eta=0.16$.

The above properties are found to decrease monotonically with
increasing $\eta$ beyond $0.16$. These parameters remain
ineffective in determining the orientational order-disorder transition
that occur at $\eta\approx0.19$. However, near to the transition point
($\eta=0.20$) the largest flock sizes are $M_s\approx2000$,
$M_f\approx400$ and the extensions are $\xi_s\approx9$,
$\xi_f\approx7$. The number of flocks are $N_{c,s}\approx350$,
$N_{c,f}\approx4000$. Both the orderliness $\sigma_s$ and $\sigma_f$
are close to $\pi$. In the high noise limit ($0.40\le\eta\le0.70$),
both sSPPs and fSPPs have similar properties such as largest flock
sizes are $M_s=M_f\approx60$, extensions are
$\xi_s=\xi_f\approx4$. The number of flocks are
$N_{c,s}=N_{c,f}\approx10000$ when total number sSPPs and fSPPs are
$N_f=N_s=16384$. The orderliness of both the species is the lowest,
$\sigma_f=\sigma_s=2\pi$.

\begin{figure}[t]
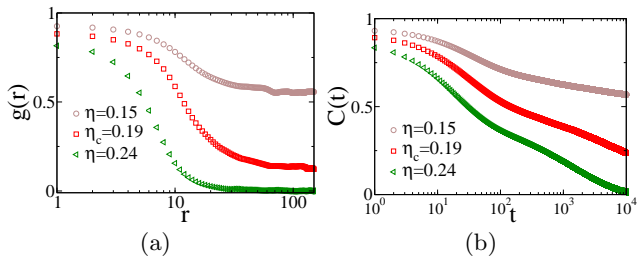

  \centerline{
         \hfill\psfig{file=fig5a.eps,width=0.23\textwidth}
     \hfill\psfig{file=fig5b.eps,width=0.23\textwidth}
  \hfill }
   \centerline{\hfill (a) \hfill\hfill (b) \hfill}
   \caption{\label{corrln} $v_{f}=30 v_s, v_s=0.01$: (a) Plot of $g(r)$
     against $r$. (b) Plot of $C(t)$ against $t$. System size is
     $L=256$.} 
\end{figure} 

\subsection{Velocity correlations}
The spatial correlation is measured by
the two-point velocity correlation function $g(r)$ and the temporal
correlation is measured by the velocity time auto-correlation function
$C(t)$. Following Ref.\cite{romenskyy2013statistical}, they are given
by
 \begin{equation}
   \label{t_corr}
     g(r)= \left \langle \frac{\vec{v}_{i}(0)
  \cdot \vec{v}_{j}(r)}{|\vec{v}_{i}(0)|
   |\vec{v}_{j}(r)|}\right \rangle,
  C(t)=\left \langle \frac{\vec{v}_{i}(0)
  \cdot \vec{v}_{i}(t)}{|\vec{v}_{i}(0)|
    |\vec{v}_{i}(t)|}\right \rangle
 \end{equation}
where $i$ and $j$ are particle indices, $t$ is the time interval, $r$
is distance between $i$ and $j$ and $\langle\cdots\rangle$ stands for
an ensemble average over $1000\times N$ configurations. The values of
$g(r)$ versus $r$ and $C(t)$ versus $t$ are estimated for three
different noise values around the critical point as $\eta=0.15$,
$0.19(\eta_c)$ and $0.24$. The $g(r)$ and $C(t)$ values are plotted in
Fig.\ref{corrln}(a) and (b) against $r$ and $t$ respectively in
semi-logarithmic scale. The critical correlations $g(r)$ or $C(t)$ at
$\eta_c=0.19$ curves are in red, the brown curves represent the $g(r)$
or $C(t)$ in the ordered (DS+DC) phase and the green curves represent
the $g(r)$ or $C(t)$ in the disordered (RS+RC) phase.

It can be seen from Fig.\ref{corrln}(a), the correlation $g(r)$
decays faster for $r>10$. In this region, the extension of the flocks
($\xi_p$), is found to be of the order of $\xi_f,\xi_s\approx 10$.
As a result, the velocity orientation is correlated among the SPPs present
in the same flock, and the correlation is found to be strong for $r<10$.
In the sub-critical regime ($\eta<0.19$, shown in brown), a strong correlation
($g(r)>0.5$) sustains all over the system. Whereas, in the upper critical
regime ($\eta>0.19$, shown in green), the correlation decays very fast and
goes to zero at $r\approx 50$. However, the critical correlation $g(r)$
($\eta_c=0.19$) sustains with a smaller value ($g(r)\approx0.15$) up to a
distance of the order of system size. It is expected that, at the criticality,
the correlation length should be of the order of system size. With a
small increase in noise, such correlation could be destroyed, as seen in the
case of $\eta=0.24$, and reducing the noise, $g(r)$ increases as it is seen
in the case of $\eta=0.15$. It is important to notice that even though the
interaction is short-ranged, the correlation extends up to system size.

It can be seen in Fig.\ref{corrln}(b) that the auto-correlation
$C(t)$ is very high over a long time $t$, irrespective of $\eta$.
The critical $C(t)$ sustains for a long time $t>10^{4}$. The
$C(t)$ at $\eta_c$ is always less than $C(t)$ for the DS+DC phase, and it is
always greater than $C(t)$ for the RS+RC as expected. The $C(t)$ for
$\eta=0.24$ goes to zero at $t=10^4$ whereas for $\eta=0.15$, it is
$\approx0.56$ at $t=10^4$ continues to be high beyond
this $t$. It seems that once dense flocks are formed in the system,
they stay for a long time, and a strong correlation exists in the ordered phase,
whereas the flocks in the disordered phase seem to be less stable due to
the randomness among the SPPs.

\section{Phase transition}
As the angular noise $\eta$ varies, the system undergoes a transition
from an orientational ordered state to an orientational disordered
state. In this model, such a transition occurs when the DS+DC phase
(Fig.\ref{morph_all}(b)) changes to the RS+RC phase
(Fig.\ref{morph_all}(c)) around the critical noise $\eta=\eta_c$. The order
parameter of the transition is defined as
\begin{equation}
  \label{d_opw}
\phi(\eta,L)= \sum_{p=f,s}\phi_p(\eta,L), \ \phi_p(\eta,L)=
\frac{1}{N_p}\left|\sum_{i=1}^{N_p}\frac{{\vec v}_{p,i}} {|{\vec
    v}_{p,i}|}\right|
\end{equation}
where $\phi$ represents the total order-parameter, $\phi_p$
($p\in\{s,f\}$) is the partial order-parameter and $N_p$ is the number
of $p$-type SPPs. The susceptibility $\chi$ for the whole system and
that of the partial systems $\chi_p$ can be estimated from the
fluctuation in their respective order parameters $\phi$ and $\phi_p$
as
\begin{equation}
  \label{d_susc}
  \chi= L^2 \left[\langle\phi^2\rangle - \langle\phi\rangle^2\right],
  \ \ \ \chi_p = L^2 \left[\langle\phi_{p}^2\rangle -
    \langle\phi_{p}\rangle^2\right]
\end{equation}  
where $\langle\phi^n\rangle=\int\phi^nP(\phi)d\phi$,
$\langle\phi_p^n\rangle=\int\phi_p^nP(\phi_p)d\phi_p$, $P(\phi)$ and
$P(\phi_p)$ are the distribution functions of $\phi$ and $\phi_p$
respectively. Similarly,
the fourth order Binder cumulant for the whole system and that of the
partial systems are defined as,
\begin{equation}
  \label{d_binder}
U= 1 - \frac{\langle\phi^4\rangle}{3\langle\phi^2\rangle^2}, \ \ \ U_p
= 1 - \frac{\langle\phi_{p}^4\rangle}{3\langle\phi_{p}^2\rangle^2}
\end{equation}
where the higher order averages are obtained following the definitions
of $\langle\phi^n\rangle$ and $\langle\phi_p^n\rangle$ given above.

If the orientational order-disorder transition is continuous, the finite
size scaling (FSS) relations of the above parameters can be given
following the equilibrium thermal critical phenomena
\cite{binder1987theory,christensen2005complexity}, as
\begin{equation}
  \label{d7}
   \phi(\eta,L)=L^{-\beta/\nu}\phi_{0}[\epsilon L^{1/\nu}]
\end{equation} 
where $\epsilon=(\eta-\eta_c)/\eta_c$ the reduced noise, $\beta$ is
the order parameter exponent, $\nu$ is the correlation length exponent
and $\phi_{0}$ is a scaling function. At the criticality
$\eta=\eta_c$, $\phi(\eta_c,L)\sim L^{-\beta/\nu}$. The order
parameter distribution $P_L(\phi)$ for a given system of size $L$ is
defined as
\begin{equation}
  \label{pd}
  P_L(\phi) = L^{\beta/\nu} \widetilde{P}_L\left[\phi L^{\beta/\nu}\right]
\end{equation} 
where $\widetilde{P}_L$ is a scaling function. At the criticality, the
distribution $P_L(\phi)$ is unimodal for a continuous transition. The
FSS form of the susceptibility is given by
\begin{equation}
  \label{d8}
  \chi(\eta,L)= L^{\gamma/\nu}\chi_{0}[\epsilon L^{1/\nu}]
\end{equation}  
where $\chi_0$ is a scaling function, $\gamma/\nu=d-2\beta/\nu$ and
$d$ ($=2$) is the space-dimension. At $\eta=\eta_c$,
$\chi(\eta_c,L)\sim L^{\gamma/\nu}$. The FSS form of the fourth order
Binder cumulant is given by
\begin{equation}
  \label{d9}
  U(\eta,L) = U_{0}[\epsilon L^{1/\nu}]
\end{equation}
where $U_{0}$ is a scaling function. The derivative of $U(\eta,L)$
with respect to $\eta$ follows a scaling relation
\cite{cambui2016critical},
\begin{equation}
  \label{d10}
  U^{\prime}(\eta,L)=L^{1/\nu}\frac{U_{0}^\prime[\epsilon L^{1/\nu}]}{\eta_{c}}
\end{equation}
where the primes on $U$ and $U_0$ denote their derivatives with respect
to $\eta$. For a continuous transition, the cumulant $U$ always remain
positive. At $\eta=\eta_c$, the cumulants of different systems of size
($L$) become independent of $L$ and $U^{\prime}(\eta_c,L)\sim L^{1/\nu}$ at
the transition.

In case the orientational order-disorder transition is discontinuous,
the order parameter exponent $\beta$ should go to zero. As a
consequence, the susceptibility should then scale as $\chi\sim L^d$,
where $d$ is the space dimension. The Binder cumulant $U$ would
exhibit a sharp fall towards a negative value at the transition
point. As the system exhibits the coexistence of two phases, the order
parameter distribution $P(\phi)$ would be a bimodal distribution.

\begin{figure}[t]
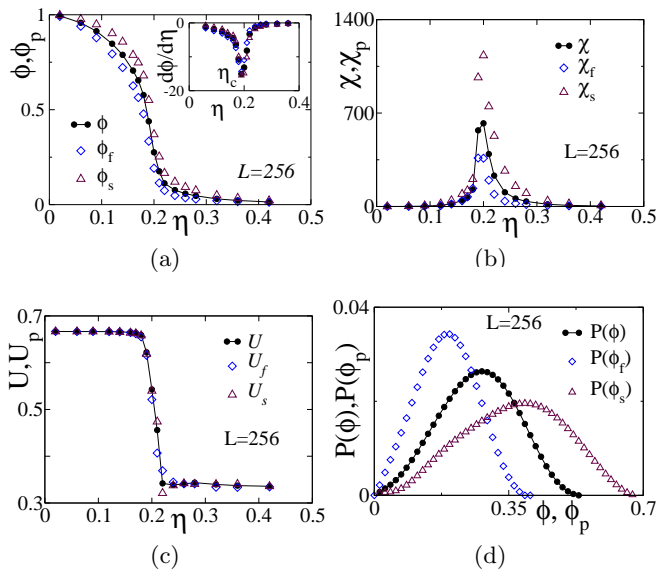

\centerline{
  \hfill\psfig{file=fig6a.eps,width=0.235\textwidth}
         \hfill\psfig{file=fig6b.eps,width=0.24\textwidth}
  \hfill }
\centerline{\hfill (a) \hfill\hfill (b) \hfill}
   \vspace{1.0 em}
   \centerline{
   \hfill\psfig{file=fig6c.eps,width=0.235\textwidth}
   \hfill\psfig{file=fig6d.eps,width=0.24\textwidth}
       \hfill }
   \centerline{\hfill (c) \hfill\hfill (d) \hfill}
   \caption{\label{op_30_256} $v_{f}=30 v_s, v_s=0.01$: (a) Plot of
     $\phi$ and $\phi_p$ vs $\eta$. Derivatives of
     $\phi$ and $\phi_p$ with respect to $\eta$ are shown in the
     inset. (b) Plot of $\chi$ and $\chi_p$ vs $\eta$.
     (c) Plot of $U$ and $U_p$ vs $\eta$. (d) Plot of $P(\phi)$
     and $P(\phi_p)$ at $\eta=\eta_c$. System size is $L=256$. }
\end{figure}

Below we discuss results of three different velocity
scenarios. Keeping $v_s=0.01$ fixed, $v_f=30v_s$;
$v_f=50v_s$; and $v_f=150v_s$ are taken. In all three scenarios, we study
variations of order parameter $\phi$, Binder cumulant $U$,
susceptibility $\chi$, and order parameter distribution $P(\phi)$
as a function of the angular noise $\eta$ on systems of
different sizes $L$ and we discuss the results together in the
discussion section.

\subsection{Velocities: $v_f=30v_s$ and $v_s=0.01$}
In Fig.\ref{op_30_256}, we present data for $\phi$, $U$, $\chi$
and $P(\phi)$ for $v_f=30v_s$ on a system of size $L=256$.
The values of $\phi$ and $\phi_p$ decrease
continuously and smoothly from a positive value to zero as $\eta$
increases. The derivatives of $\phi$ and $\phi_p$ with respect to
$\eta$ are plotted in the inset of
Fig.\ref{op_30_256}(a). A minimum or dip at $\eta\approx 0.19$ is
observed in the derivatives. This is the critical noise $\eta_c=0.19$
at which the systems undergo a phase transition. It is evident from
the configurations given in Fig.\ref{morph_all}(b) and (c) that the
system is going from an ordered phase DS+DC to a disordered phase
RS+RC around $\eta_c\approx0.19$. The fluctuations in order parameters
$\chi$ and $\chi_p$ are plotted against $\eta$ in
Fig.\ref{op_30_256}(b) for $L=256$. All the fluctuations diverge at
$\eta_c\approx0.19$. The order parameter fluctuation of the sSPPs at
$\eta_c$ is much higher than that of the fSPPs. It is
similar to the observations in the VM with monodispersed SPPs.
The Binder cumulants, $U$ and $U_p$ are plotted
against $\eta$ in Fig.\ref{op_30_256}(c) for $L=256$. It can be seen
that the values of $U$, $U_f$, and $U_s$ are all positive over the
whole range of $\eta$. The distributions of order parameters
$P(\phi)$ and $P(\phi_p)$ obtained at $\eta=\eta_{c}$ are presented in
Fig.\ref{op_30_256}(d). All three distributions are found to be
unimodal. The positive value of the Binder cumulant and unimodal
distribution of $\phi$ indicates a continuous transition.

\begin{figure}[t]
\centerline{
  \hfill\psfig{file=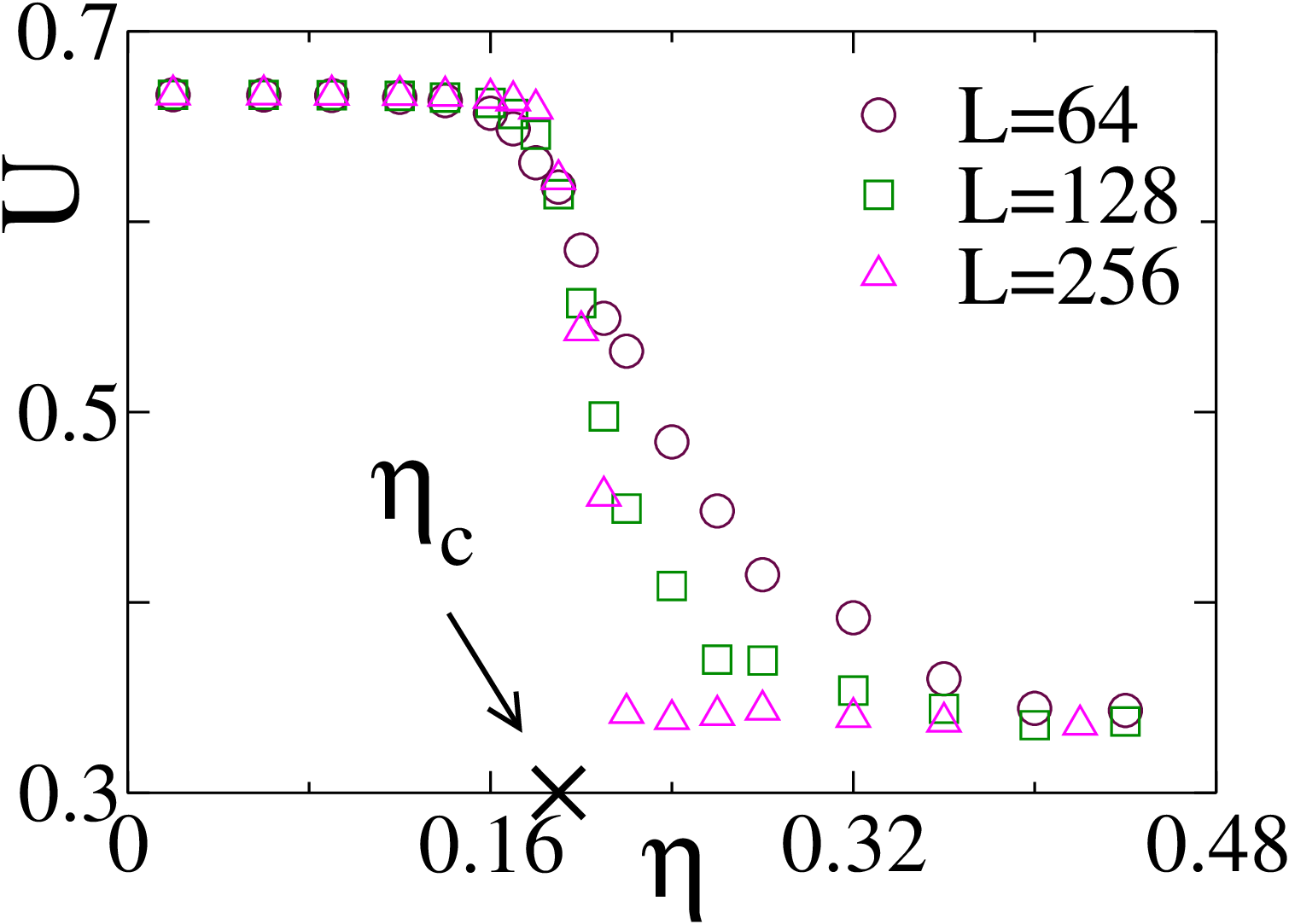,width=0.16\textwidth}
     \hfill\psfig{file=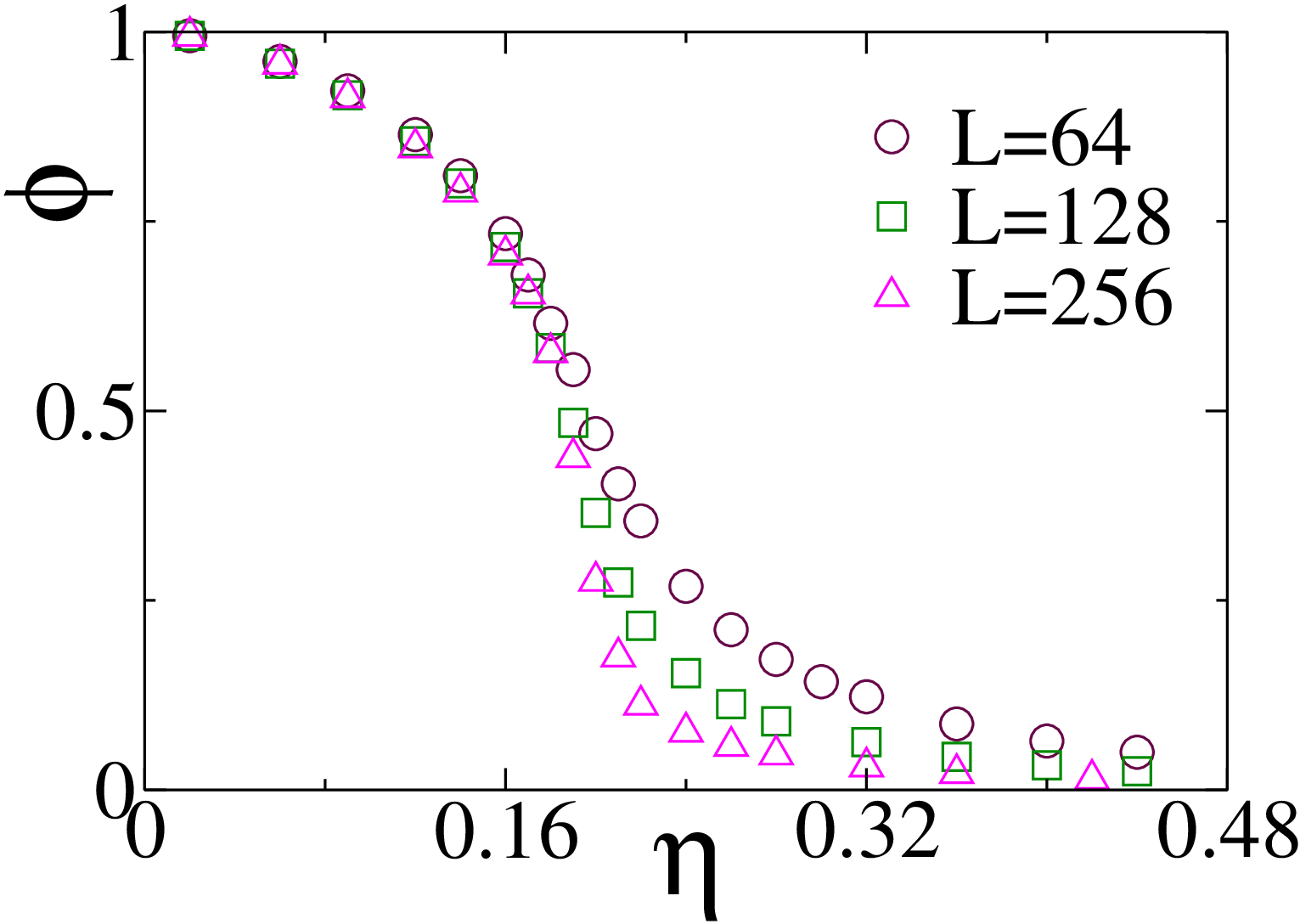,width=0.16\textwidth}
   \hfill\psfig{file=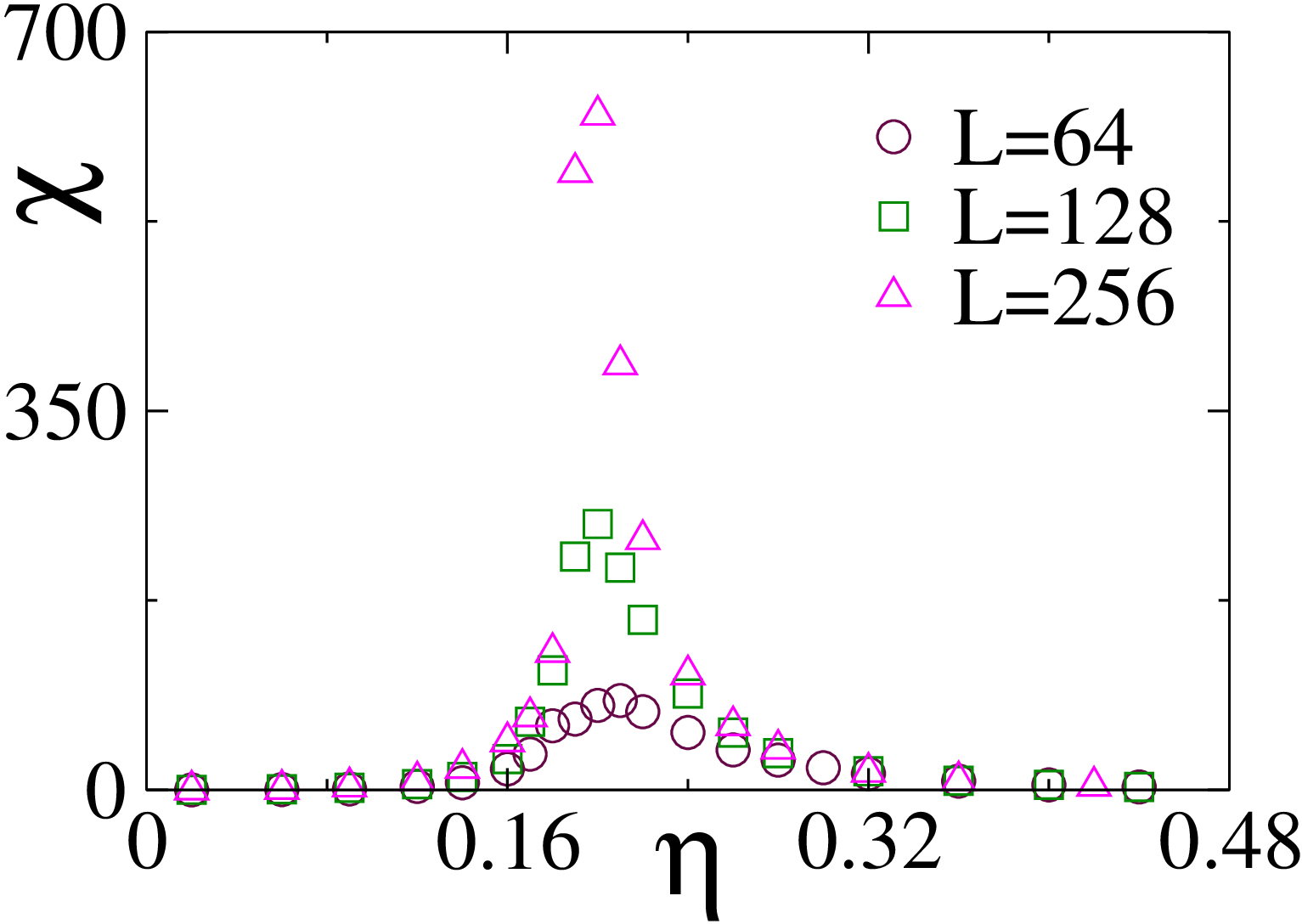,width=0.16\textwidth}
  \hfill }
\centerline{\hfill (a) \hfill\hfill (b) \hfill\hfill (c) \hfill}
   \vspace{1.0 em}
   \centerline{
       \hfill\psfig{file=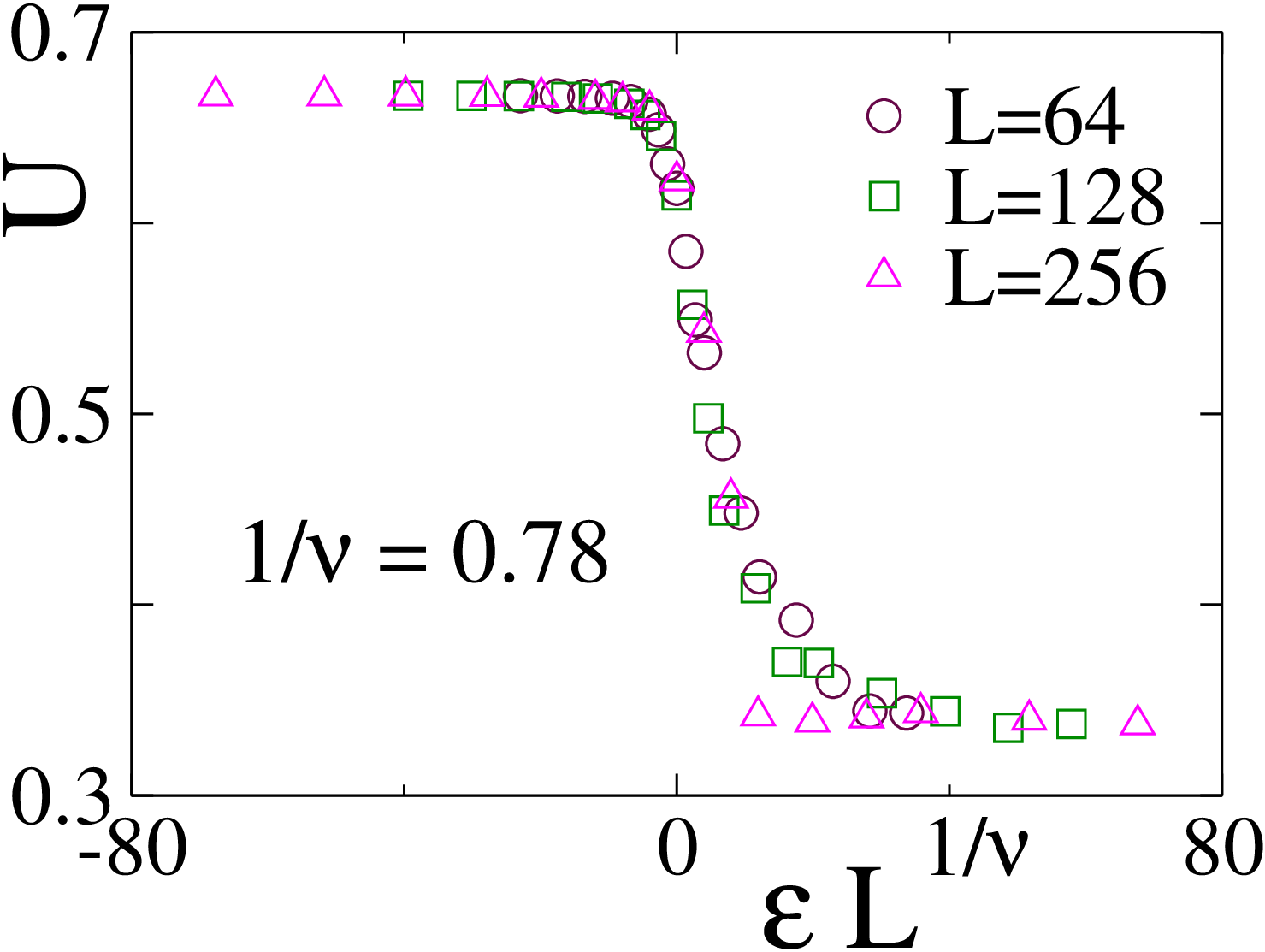,width=0.16\textwidth}
       \hfill\psfig{file=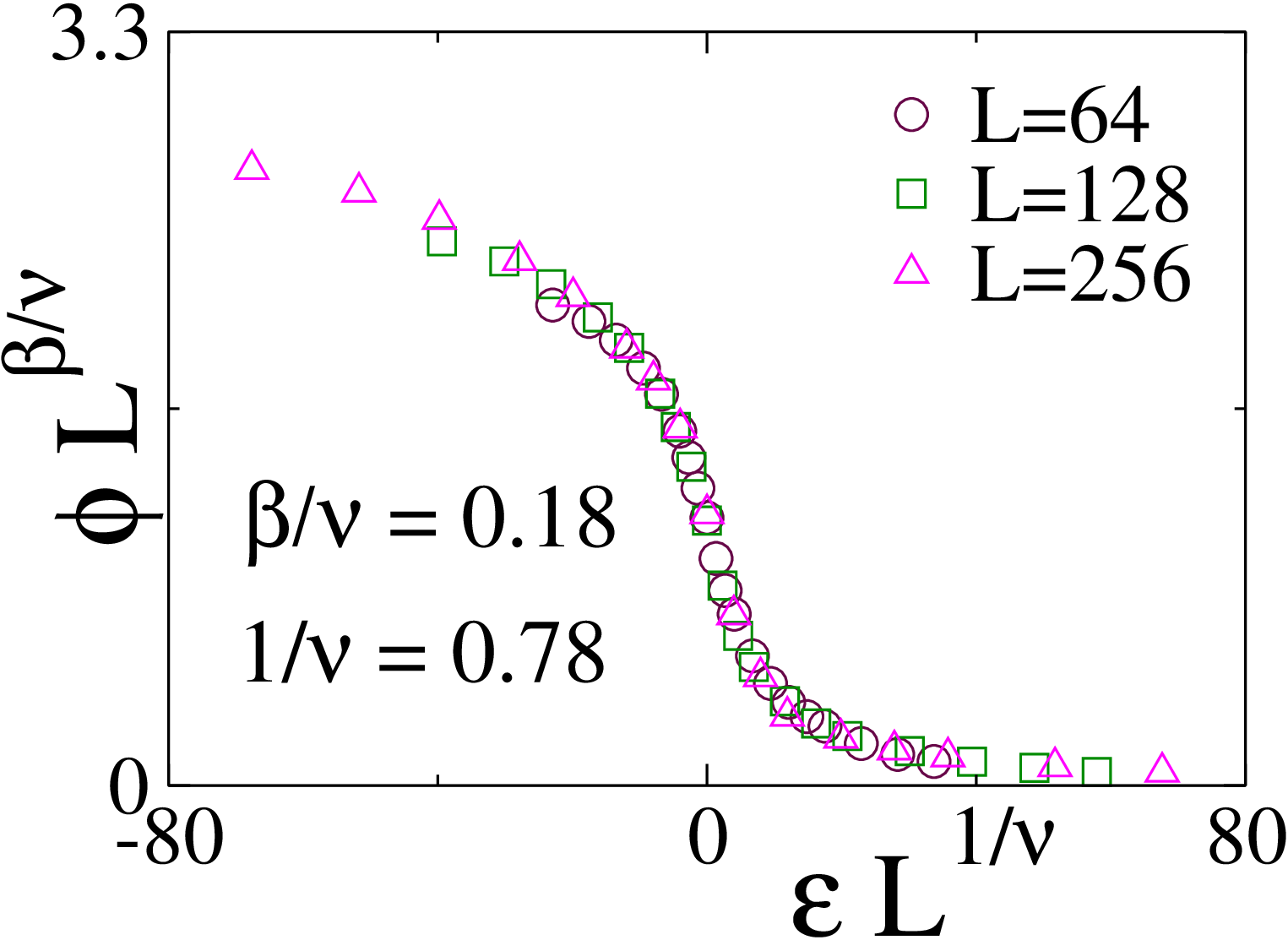,width=0.16\textwidth}
      \hfill\psfig{file=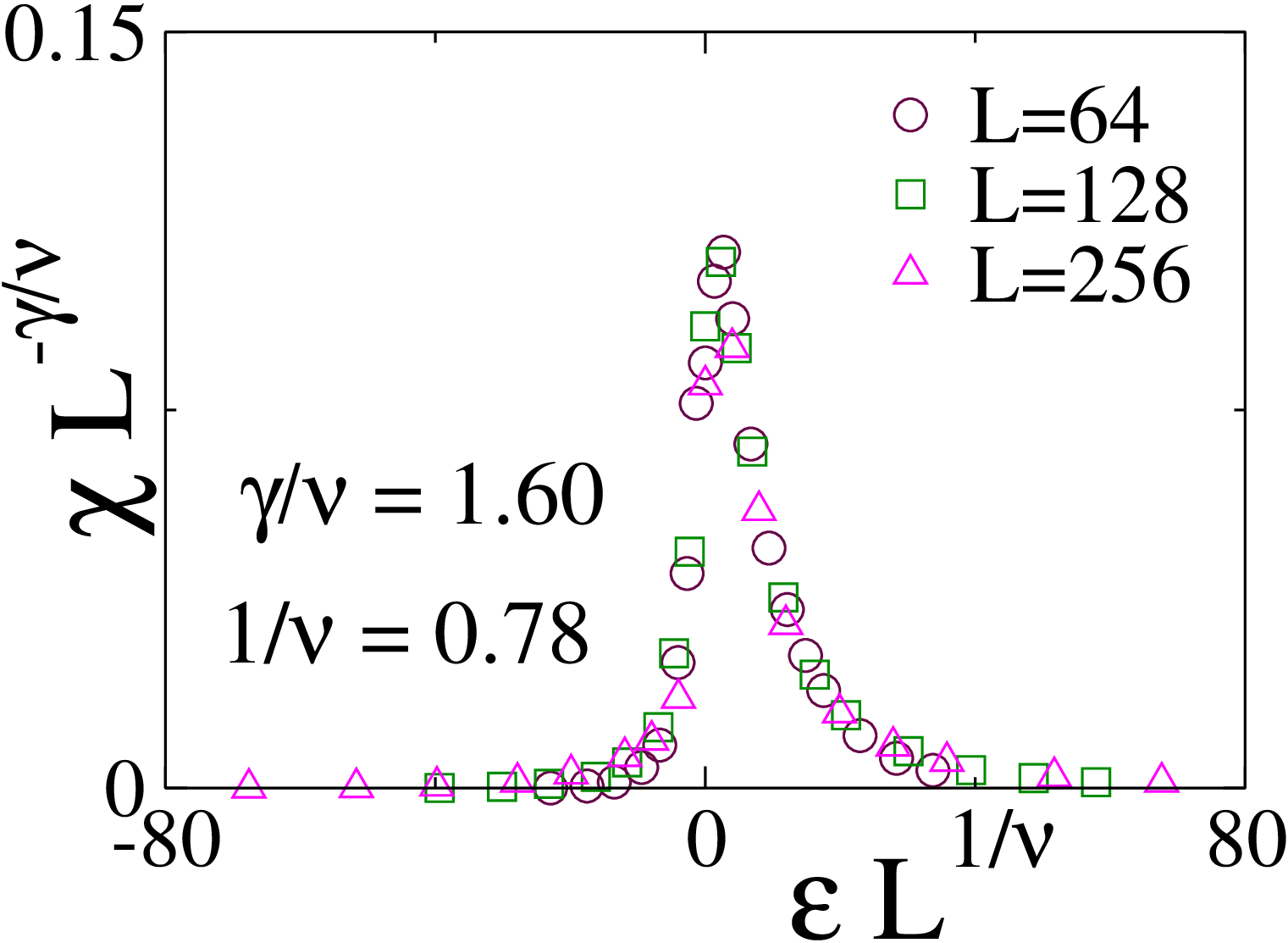,width=0.16\textwidth}
   \hfill }
\centerline{\hfill (d) \hfill\hfill (e) \hfill\hfill (f) \hfill}
 \caption{\label{fss_all_30} $v_{f}=30 v_s, v_s=0.01$: (a) Plot of $U$
   vs $\eta$, (b) plot of $\phi$ vs $\eta$ and (c) plot of
   $\chi$ vs $\eta$ for $L=64$, $128$ and $256$. The cross on the
   $\eta$-axis indicates $\eta_c$. (d) Plot of $U$ vs the scaled
   noise $\epsilon L^{1/\nu}$. (e) Plot of $\phi L^{\beta/\nu}$
   against $\epsilon L^{1/\nu}$. (f) Plot of $\chi L^{-\gamma/\nu}$
   against $\epsilon L^{1/\nu}$. The values of the exponents are taken
   as $\beta/\nu=0.18$, $\gamma/\nu=1.60$ and $1/\nu=0.78$. }
\end{figure}

The critical exponents are extracted for the whole system with
$v_f=30v_s$ performing FSS analysis. Binder cumulant $U$, order
parameter $\phi$ and susceptibility $\chi$ are plotted against the
angular noise $\eta$ for three different systems of sizes $L=64$,
$128$ and $256$ in Fig.\ref{fss_all_30}(a),(b) and (c)
respectively. The plots of $U$ versus $\eta$ (Fig.\ref{fss_all_30}(a))
for different $L$ intersect at $\eta_c\approx 0.19$, the $L$
independent critical point as expected in a continuous transition. It
is marked by a cross on the $\eta$-axis. The corresponding $U_c$ is
identified as $U_c\approx0.61$, which is close to that of a
mono-dispersed system\cite{baglietto2008finite}. A rough estimate of
the exponents $1/\nu$, $\beta/\nu$ and $\gamma/\nu$ are obtained from
the scaling relations $U^{\prime}(\eta_c,L)\sim L^{1/\nu}$,
$\phi(\eta_c,L)\sim L^{-\beta/\nu}$ and $\chi(\eta_c,L)\sim
L^{\gamma/\nu}$ at the criticality. The best possible FSS forms of the
scaled parameters against the scaled noise $\epsilon L^{1/\nu}$ are
obtained, tuning these exponents further. $U$, $\phi L^{\beta/\nu}$,
and $\chi L^{-\gamma/\nu}$ are plotted against $\epsilon L^{1/\nu}$ in
Fig.\ref{fss_all_30}(d), (e) and (f) respectively. A reasonable
collapse of data in all three cases are obtained taking $1/\nu=0.78$,
$\beta/\nu=0.18$ and $\gamma/\nu=1.60$ at $\eta_c=0.19$. The critical
exponents satisfy the scaling relation $\gamma/\nu+2\beta/\nu=2$
within error bars. The values of these critical exponents are reported
in Table.\ref{t1}. The FSS forms of $U_p$, $\phi_p$, and $\chi_p$ for
the partial systems are also verified, and the scaling relations are
satisfied with the same critical exponents within error bars. 

\subsection{Velocities: $v_f=50v_s$ and $v_s=0.01$}
\begin{figure}[t]
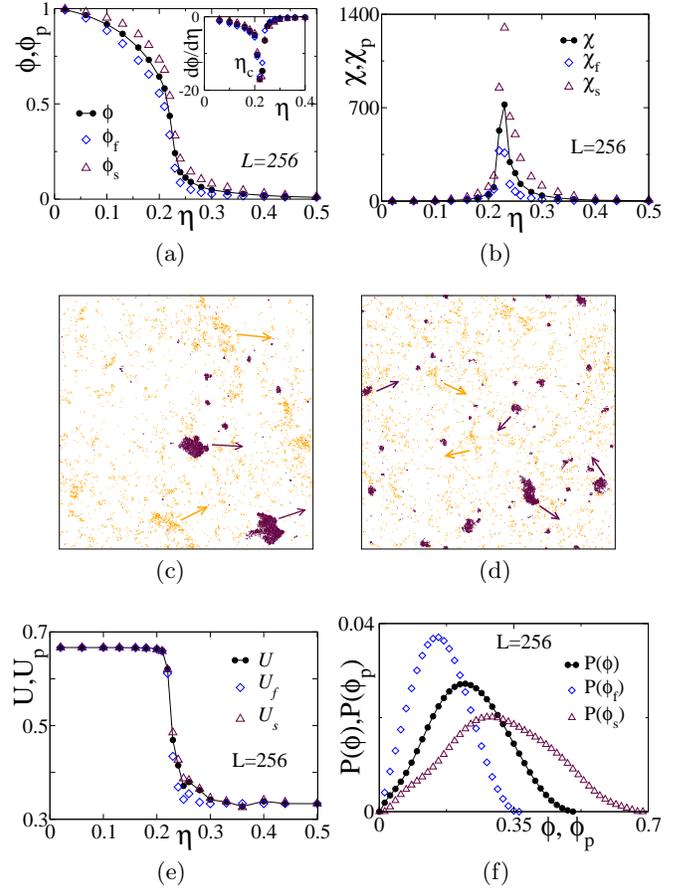

\centerline{
  \hfill\psfig{file=fig8a.eps,width=0.235\textwidth}
         \hfill\psfig{file=fig8b.eps,width=0.24\textwidth}
  \hfill }
\centerline{\hfill (a) \hfill\hfill (b) \hfill}
\vspace{1.0 em}
\centerline{
  \hfill\psfig{file=fig8c.eps,width=0.19\textwidth}
         \hfill\psfig{file=fig8d.eps,width=0.19\textwidth}
  \hfill }
\centerline{\hfill (c) \hfill\hfill (d) \hfill}
\vspace{1.2 em}
   \centerline{
   \hfill\psfig{file=fig8e.eps,width=0.235\textwidth}
   \hfill\psfig{file=fig8f.eps,width=0.24\textwidth}
       \hfill }
   \centerline{\hfill (e) \hfill\hfill (f) \hfill}
   \caption{\label{op_50_256} $v_{f}=50 v_s, v_s=0.01$: (a) Plot of
     $\phi$ and $\phi_p$ vs $\eta$. Derivatives of $\phi$ and
     $\phi_p$ with respect to $\eta$ are shown in the inset. (b) Plot
     of $\chi$ and $\chi_p$ vs $\eta$. System size is $L=256$.
     (c) Morphology of the system for $L=128$ \& $\eta=0.18$, (d)
     $\eta=0.26$. Orange: fSPPS and maroon: sSPPs. The arrows represent
     the directions of motion. (e) Plot of $U$ and $U_p$ vs $\eta$.
     (f) Plot of $P(\phi)$ and $P(\phi_p)$ at $\eta_c$. System size is
     $L=256$. }
\end{figure}

The order parameters $\phi$ and $\phi_p$ are plotted against $\eta$ in
Fig.\ref{op_50_256}(a). The values of $\phi$ and $\phi_p$
decrease smoothly to zero as $\eta$ increases. The derivatives of
$\phi$ and $\phi_p$ with respect to $\eta$ are plotted in the inset of
Fig.\ref{op_50_256}(a) and minima of the plots at the transition noise
$\eta_c\approx 0.22$ are observed. $\phi_f$ is also less than $\phi_s$
as that of $v_{f}=30v_s$ situation. The respective fluctuations in
order parameter, $\chi$, and $\chi_p$ are plotted against $\eta$ in
Fig.\ref{op_50_256}(b). They diverge at $\eta_c \approx 0.22$ for all
the SPPs. The morphology of the system around the transition point is
presented in Fig.\ref{op_50_256}. The phase transition occurs at
$\eta_c\approx0.22$ between the DS+DC phase and the RS+RC phase. The
DS+DC phase at $\eta=0.18$ is shown in Fig.\ref{op_50_256}(c) and the
RS+RC phase at $\eta=0.26$ is shown in Fig.\ref{op_50_256}(d). The
arrows indicate the direction of motions of the clusters and clumps of
different phases. It can be noticed that all the clusters and clumps
in DS+DC are moving in a similar direction, whereas they are random in
RS+RC.

In Fig.\ref{op_50_256}(e), $U$ and $U_p$ are plotted against $\eta$. Both
$U$ and $U_p$ remain positive over the whole range of $\eta$. In
Fig.\ref{op_50_256}(f), the distribution of order parameters $P(\phi)$
and $P(\phi_p)$ are plotted at $\eta=\eta_{c}$. The distributions are
unimodal. The positive Binder cumulants and unimodal distributions of
order parameters indicate a continuous transition in the whole system
as well as in the partial systems for the case of $v_f=50v_s$.

\begin{figure}[t]
\centerline{
  \hfill\psfig{file=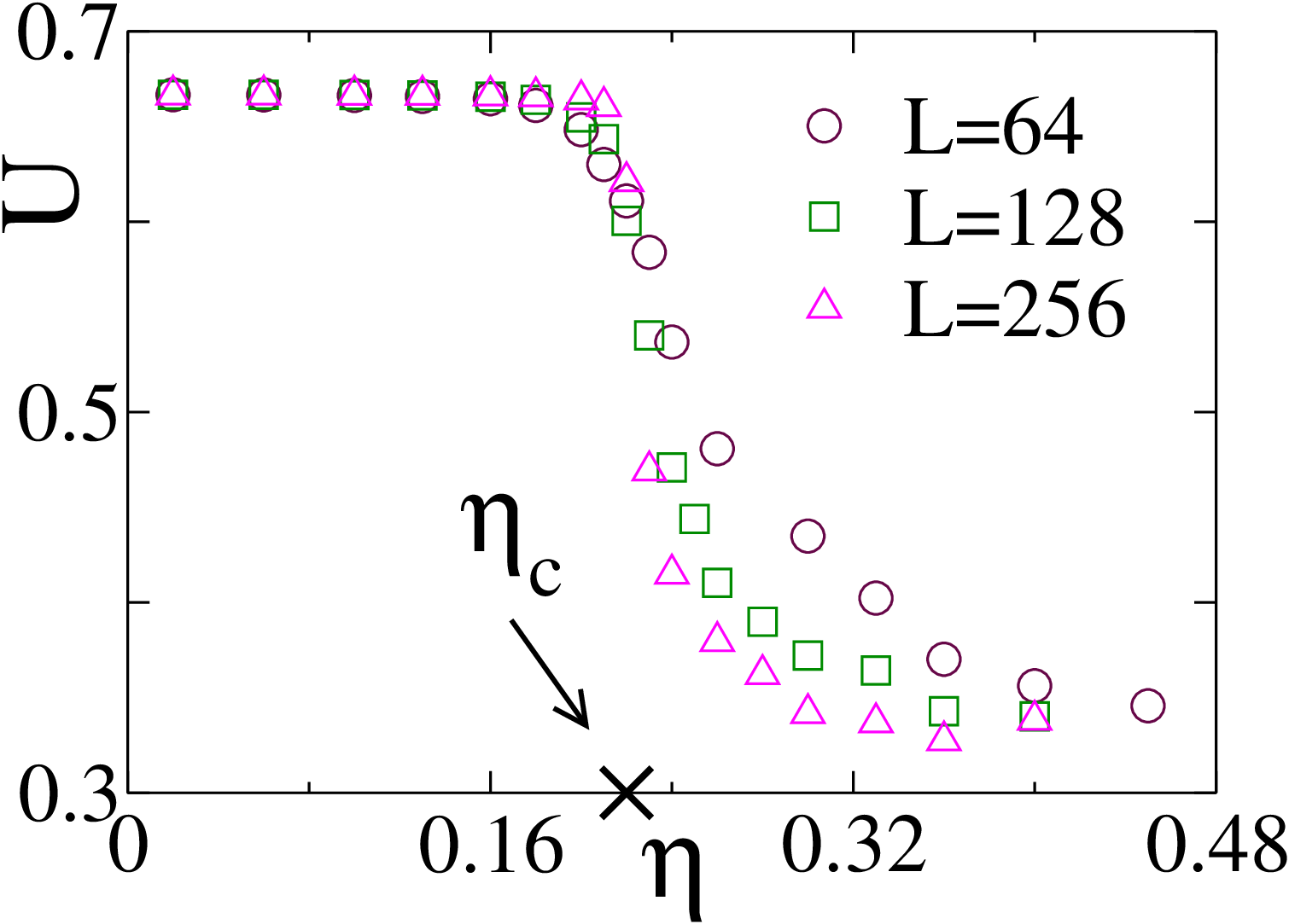,width=0.16\textwidth}
     \hfill\psfig{file=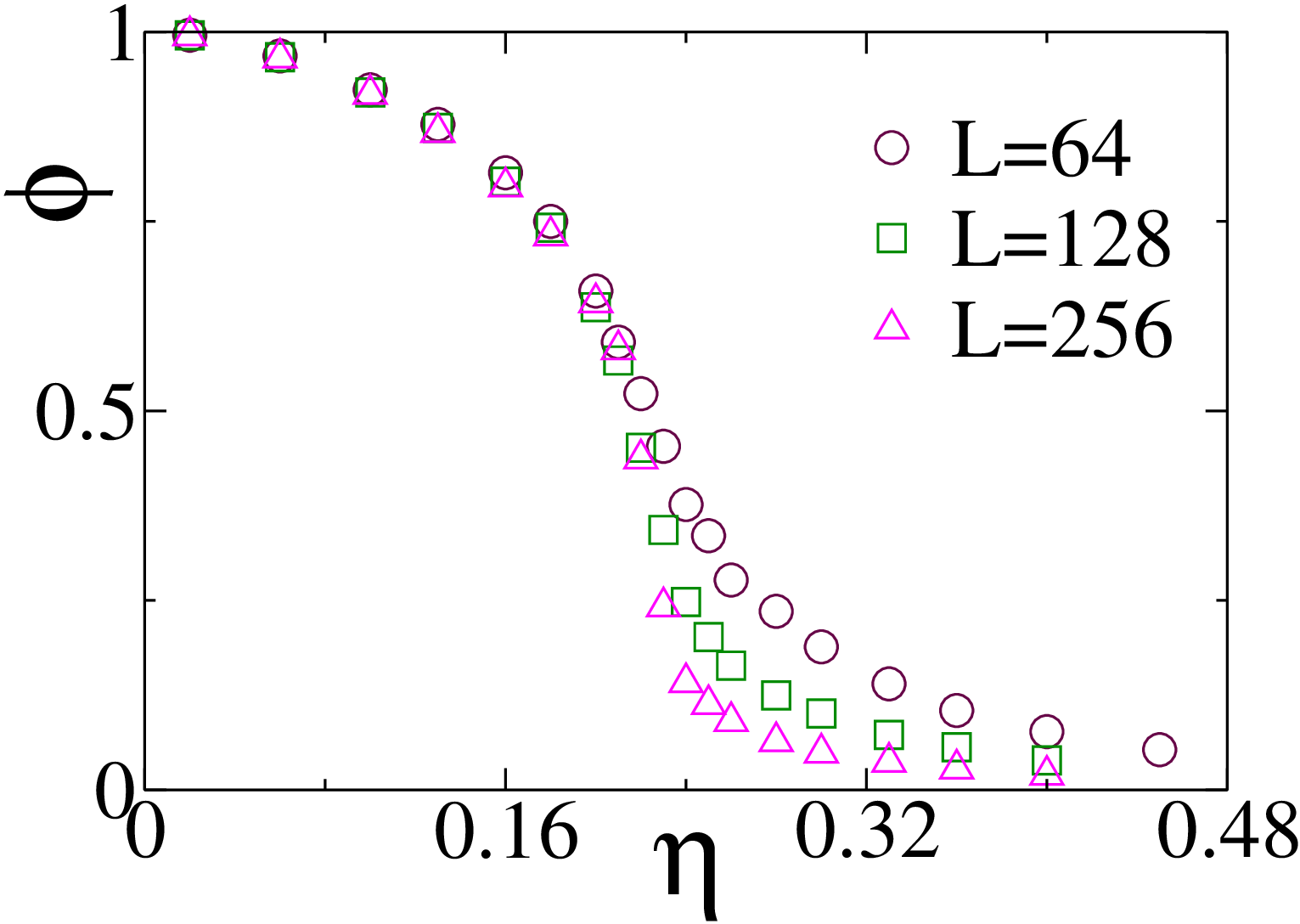,width=0.16\textwidth}
   \hfill\psfig{file=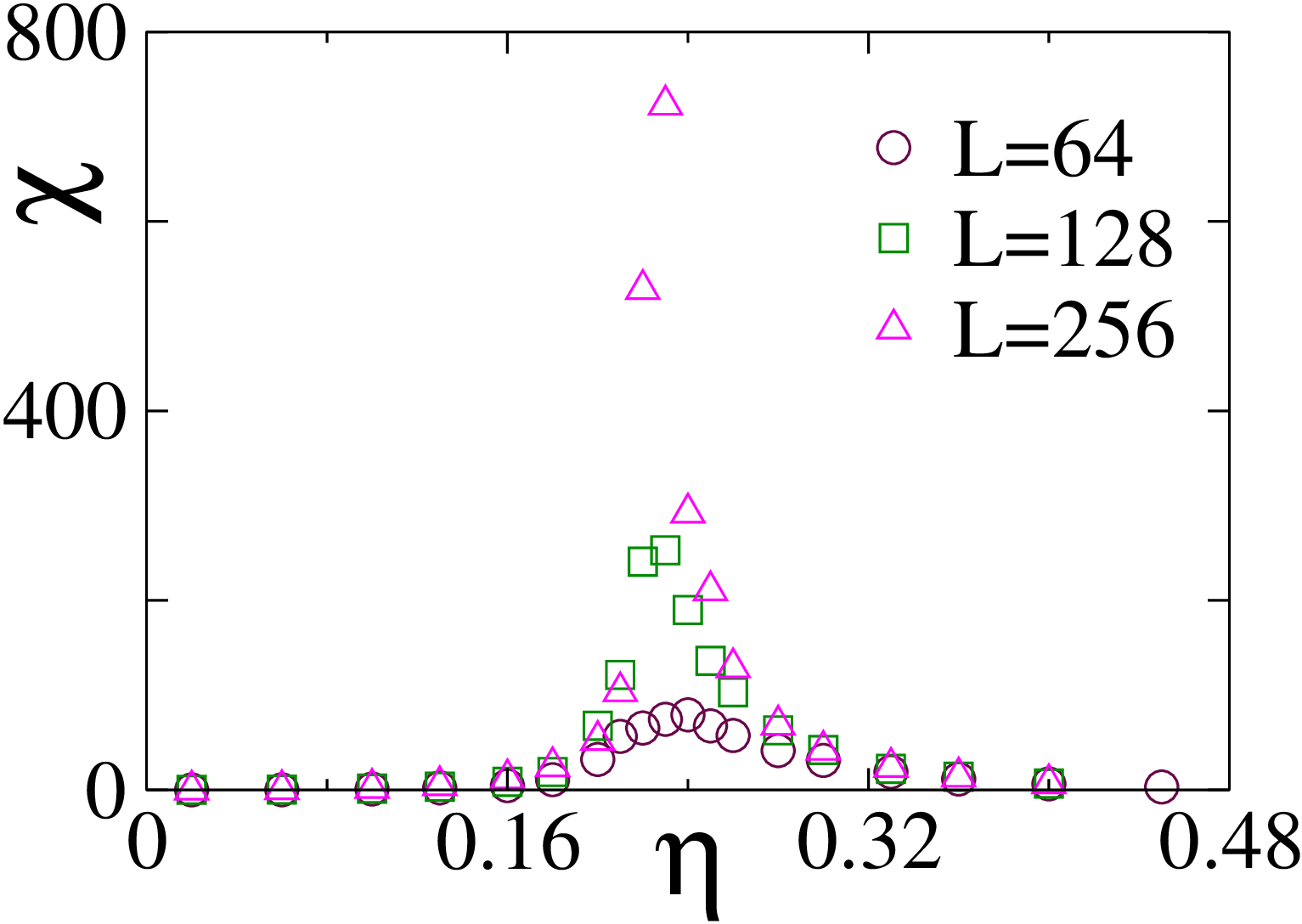,width=0.16\textwidth}
  \hfill }
\centerline{\hfill (a) \hfill\hfill (b) \hfill\hfill (c) \hfill}
   \vspace{1.0 em}
   \centerline{
       \hfill\psfig{file=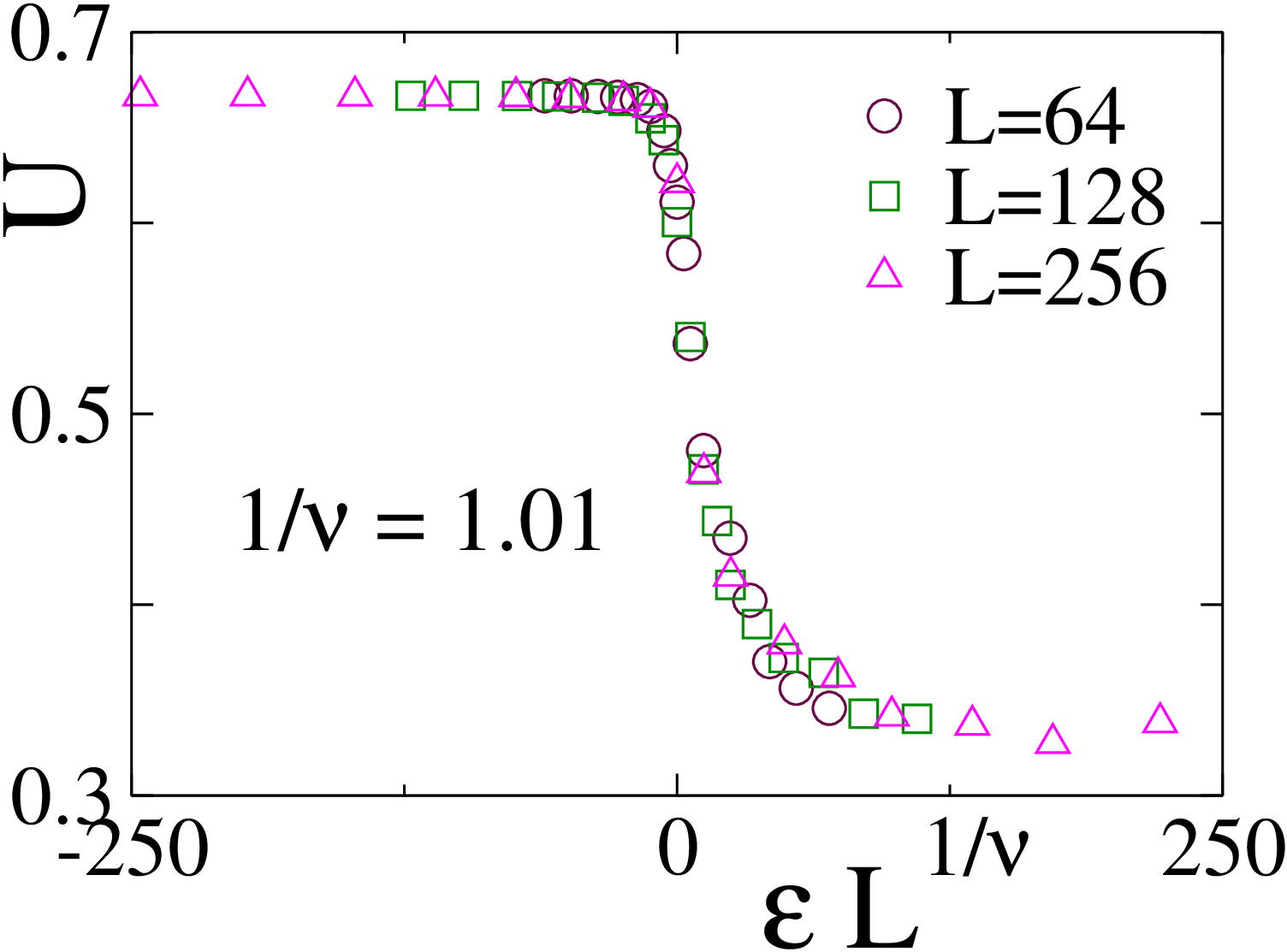,width=0.16\textwidth}
       \hfill\psfig{file=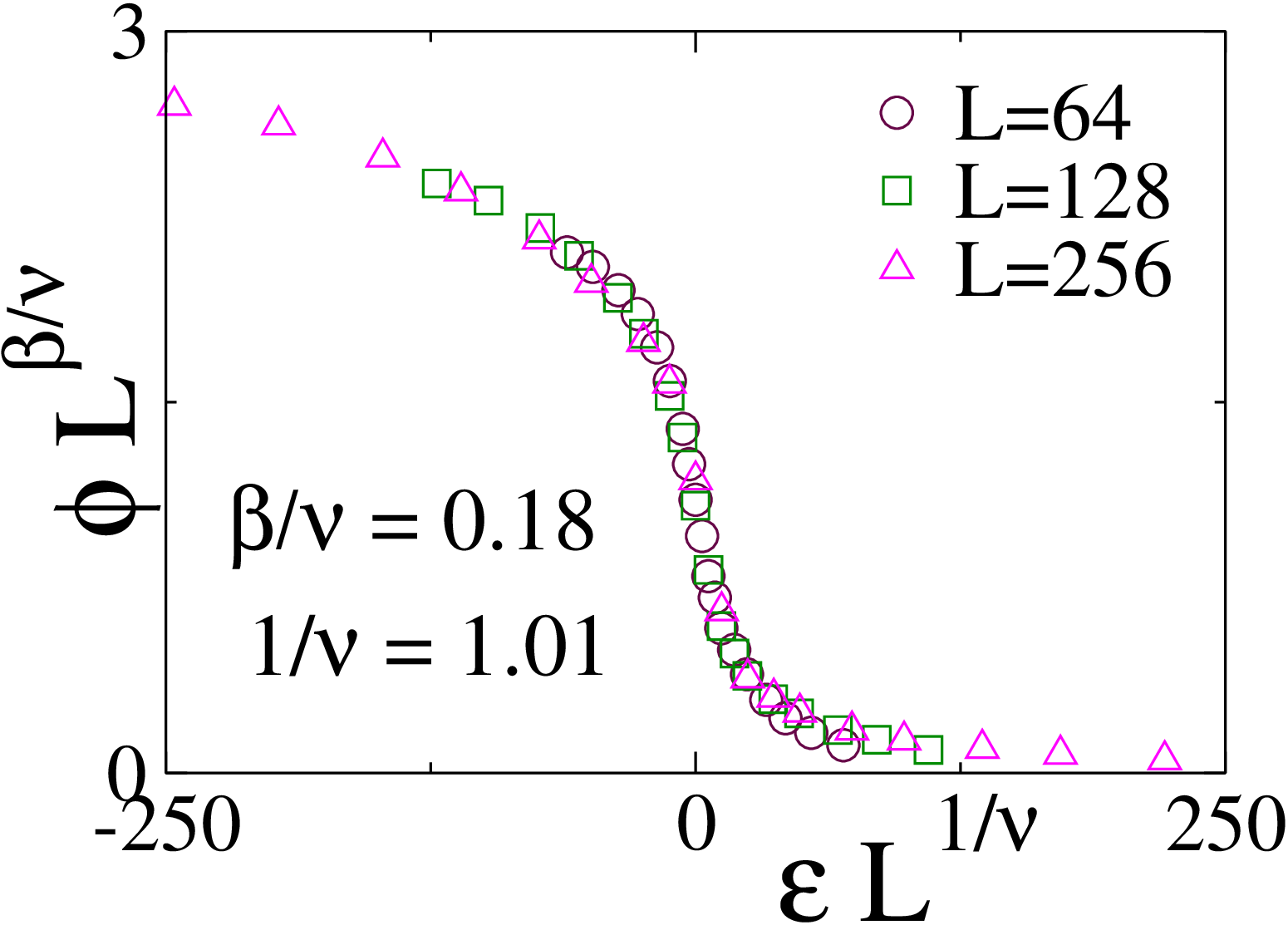,width=0.16\textwidth}
      \hfill\psfig{file=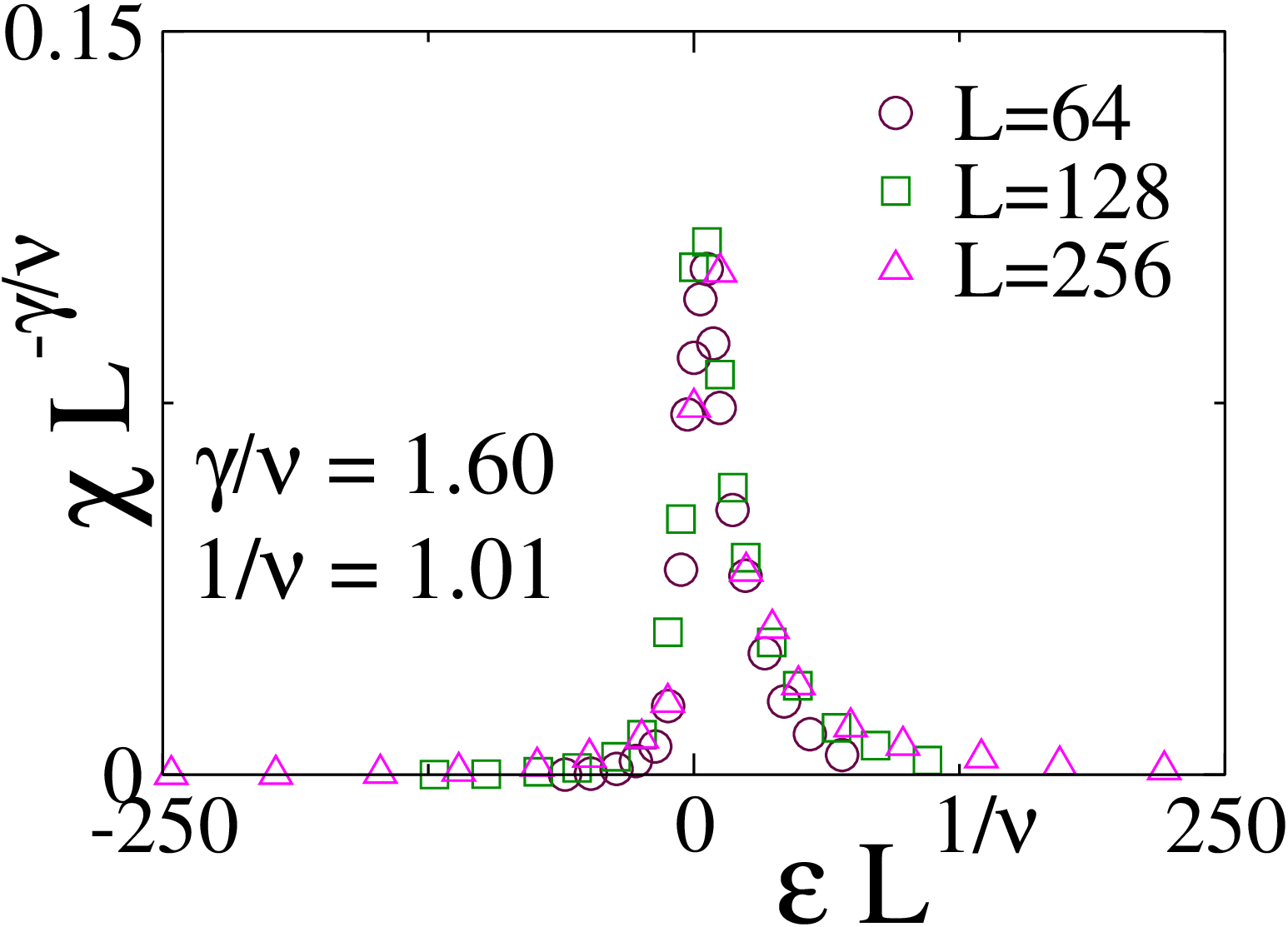,width=0.16\textwidth}
   \hfill }
\centerline{\hfill (d) \hfill\hfill (e) \hfill\hfill (f) \hfill}
\caption{\label{fss_all_50} $v_f=50 v_s, v_s=0.01$: (a) Plot of $U$
  vs $\eta$, (b) plot of $\phi$ vs $\eta$ and (c) plot of
  $\chi$ vs $\eta$ for $L=64$, $128$ and $256$. The cross on the
  $\eta$-axis indicates $\eta_c$. (d) Plot of $U$ vs the scaled
  noise $\epsilon L^{1/\nu}$. (e) Plot of $\phi L^{\beta/\nu}$ against
  $\epsilon L^{1/\nu}$. (f) Plot of $\chi L^{-\gamma/\nu}$ against
  $\epsilon L^{1/\nu}$. The values of the exponents are taken as
  $\beta/\nu=0.18$, $\gamma/\nu=1.60$ and $1/\nu=1.01$. }
\end{figure}

Since the transitions are continuous, we perform the FSS analysis to
extract the critical exponents. The values of $U$, $\phi$ and $\chi$
are plotted against $\eta$ in Fig.\ref{fss_all_50}(a), (b) and (c)
respectively for different system sizes $L=64$, $128$ and $256$. The
plots of $U$ versus $\eta$ in Fig.\ref{fss_all_50}(a) for different
$L$ intersect at $\eta_c\approx 0.22$, marked by a cross on the
$\eta$-axis.  The corresponding critical value of the cumulant is
$U_c\approx 0.61$. It seems the value of $U_c$ is independent of the
velocity ratio in this model. As followed in the case of $v_f=30v_s$,
rough estimates of the exponents $1/\nu$, $\beta/\nu$ and $\gamma/\nu$
are obtained from the critical scaling relations
$U^{\prime}(\eta_c,L)\sim L^{1/\nu}$, $\phi(\eta_c,L)\sim
L^{-\beta/\nu}$ and $\chi(\eta_c,L)\sim L^{\gamma/\nu}$. The best
possible FSS form of the scaled parameters against the scaled noise
$\epsilon L^{1/\nu}$ are obtained tuning these exponents further. The
values of $U$, the scaled order parameter $\phi L^{\beta/\nu}$, and
the scaled susceptibility $\chi L^{-\gamma/\nu}$ are plotted against
$\epsilon L^{1/\nu}$ in Fig.\ref{fss_all_50}(d), (e) and (f)
respectively. A reasonable collapse of data in all three cases are
obtained tuning the exponents further to $1/\nu=1.01$,
$\beta/\nu=0.18$ and $\gamma/\nu=1.60$. The exponents satisfy the
scaling relation $\gamma/\nu+2\beta/\nu=2$ within error bars.
The values of the critical exponents obtained are also given in
Table.\ref{t1} and compared with others.

\begin{table}[h]
\begin{center}
\begin{tabular}{ |c|c|c|c| } 
\hline
Exponents & VM & BM: $v_f=30v_s$, & BM: $v_f=50v_s$,\\
& $v_0=0.1$ & $v_s=0.01$ & $v_s=0.01$\\
\hline
\multirow{1}{0.1em}{} $1/\nu$  & 0.62(12) & 0.78(9) & 1.01(5)\\ 
\hline
\multirow{1}{0.1em}{} $\beta/\nu$  & 0.275(5)  & 0.18(1) & 0.18(2)\\ 
\hline
\multirow{1}{0.1em}{} $\gamma/\nu$ & 1.45(2) & 1.60(4) & 1.60(8)\\ 
\hline
\multirow{1}{0.1em}{} $\gamma/\nu+2\beta/\nu$ & 2.00(3) & 1.96(4) & 1.96(8)\\ 
\hline
\end{tabular}
\end{center}
\caption{\label{t1} Values of the critical exponents obtained for the
  BM at different velocities of fSPPs $v_f=30v_s$ and
  $50v_s$ keeping $v_s=0.01$. The exponents for the VM with $v_0=0.1$
  and density $\rho=1/8$ to $3/4$ (taken from
  Ref.\cite{baglietto2008finite}) are presented for comparison.}
\end{table}

\subsection{Velocities: $v_f=150v_s$ and $v_s=0.01$}
\begin{figure}[t]
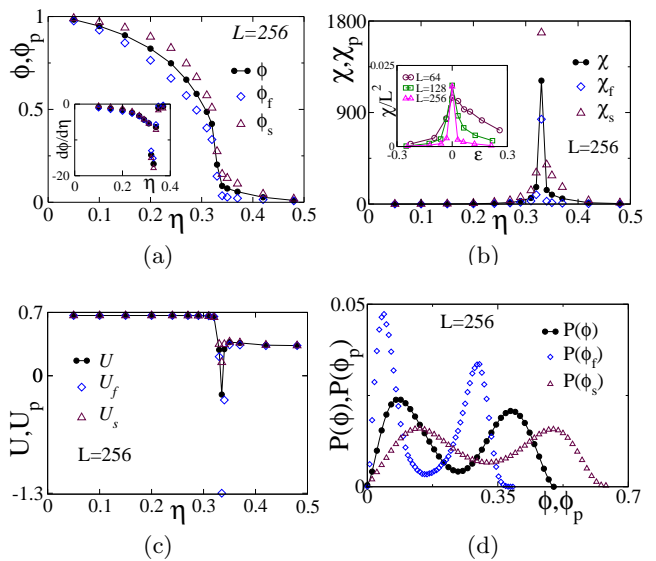

  \centerline{
    \hfill\psfig{file=fig10a.eps,width=0.23\textwidth}
       \hfill\psfig{file=fig10b.eps,width=0.235\textwidth}
  \hfill }
\centerline{\hfill (a) \hfill\hfill (b) \hfill}
   \vspace{1.0 em}
     \centerline{
       \hfill\psfig{file=fig10c.eps,width=0.23\textwidth}
       \hfill\psfig{file=fig10d.eps,width=0.232\textwidth}
        \hfill }      
\centerline{\hfill (c) \hfill\hfill (d) \hfill}
\caption{\label{op_150_256} $v_{f}=150 v_s, v_s=0.01$: (a) Plot of
  $\phi$ and $\phi_p$ vs $\eta$. Derivatives of $\phi$ and
  $\phi_p$ with respect to $\eta$ are shown in the inset. (b) Plot of
  $\chi$ and $\chi_p$ vs $\eta$. In the inset, $\chi L^{-2}$ is
  plotted against $\epsilon$ for $L=64$, $L=128$ and $L=256$. (c) Plot
  of $U$ and $U_p$ vs $\eta$. (d) Plot of $P(\phi)$ and
  $P(\phi_p)$ at $\eta_c$. }
\end{figure}

We now present simulation results obtained for $v_f=150v_s$
keeping $v_s=0.01$. The order parameters, $\phi$ and $\phi_p$ are plotted
against $\eta$ in
Fig.\ref{op_150_256}(a). There are jumps in the values of $\phi$ and
$\phi_p$ near the transition. The derivatives of $\phi$ and $\phi_p$
with respect to $\eta$ are plotted in the inset of
Fig.\ref{op_150_256}(a) and sharp minima are observed at
$\eta_c\approx 0.33$ for the whole and the partial systems. The
respective fluctuations in order parameter, $\chi$ and $\chi_p$ are
plotted against $\eta$ in Fig.\ref{op_150_256}(b). There is a sharp
peak in the fluctuations at $\eta_c\approx 0.33$ for all the SPPs. In
the inset of Fig.\ref{op_150_256}(b), the scaled fluctuations
$\chi/L^2$ are plotted against $\epsilon$ for different systems of
sizes $L=64$, $128$ and $256$. The critical values (peak values) of
scaled $\chi$ become almost $L$ independent. Hence, $\chi\sim L^2$, as
in the case of a discontinuous transition.

The Binder cumulants $U$ and $U_p$ versus $\eta$ plots are shown in
Fig.\ref{op_150_256}(c). The cumulant for the fSPPs, $U_f$
and that of the whole system $U$ have sharp negative dips at the
transition. However, the cumulant of the sSPPs ($U_s$) has
a dip at the transition, but it is yet to achieve a negative value. The
distributions of order parameters $P(\phi)$ and $P(\phi_p)$ at
$\eta=\eta_c$ are shown in Fig.\ref{op_150_256}(d). For this high velocity,
all three distributions exhibit bimodal distributions. These are characteristic
features of a discontinuous transition. Since discontinuous transition is
known to occur at high velocities in the monodispersed SPPs \cite{chate2008},
it is expected to occur in the case of fSPPs, but it is
surprising that the sSPPs are also undergoing discontinuous
transition, which is unusual for sSPPs with such a low velocity
$v_s=0.01$. At the transition, dense traveling bands of SPPs form and
disappear, resulting coexistence of two phases in the system. It is
then important to verify whether this is the case for the fSPPs here
or not. More interestingly, what will the two phases be for the sSPPs?
Below we explain the situation by studying the time evolution of the
system morphology at the transition region.

\begin{figure}[t]
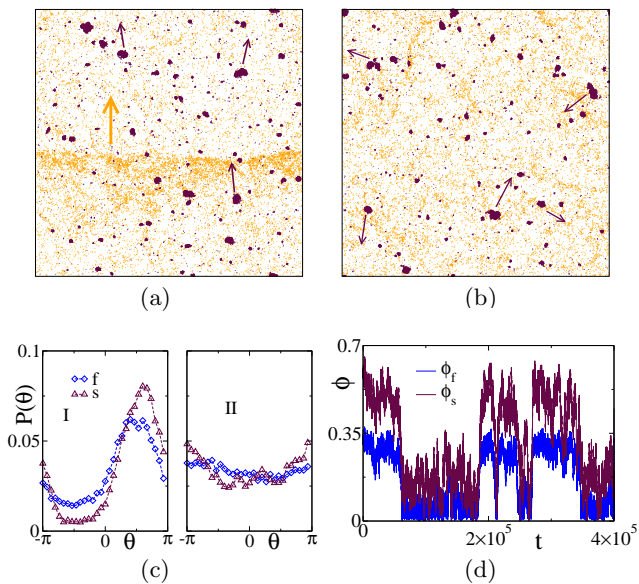

\centerline{
  \hfill\psfig{file=fig11a.eps,width=0.20\textwidth}
    \hfill\psfig{file=fig11b.eps,width=0.20\textwidth}
       \hfill }
\centerline{\hfill (a) \hfill\hfill (b) \hfill}
     \vspace{1.0 em}
\centerline{
  \hfill\psfig{file=fig11c.eps,width=0.23\textwidth}
  \hfill\psfig{file=fig11d.eps,width=0.23\textwidth}
       \hfill }
\centerline{\hfill (c) \hfill\hfill (d) \hfill}
\caption{\label{confi_150} $v_{f}=150 v_s, v_s=0.01$: Morphology at
  (a) $t=3\times 10^5$ and (b) $t=4\times 10^5$ for $\eta_c=0.33$ on a
  system of size $L=256$. Orange: fSPPs and maroon: sSPPs.
  (c) $P(\theta)$ of the morphology (a) and (b)
  are plotted in I and II. (d) Order-parameter dynamics at
  $\eta_c$. }
\end{figure}

The morphology of a system of size $L=256$ at two different time
instants $3\times 10^5$ and $4\times 10^5$ are shown in
Fig.\ref{confi_150}(a) and Fig.\ref{confi_150}(b) respectively for
$\eta=\eta_c$. In Fig.\ref{confi_150}(a), a dense traveling band of
fSPPs and clumps of sSPPs are seen. Interestingly, the band and clumps
are moving in the same direction, as indicated by the orange and maroon
arrows, respectively. However, in Fig.\ref{confi_150}(b), the dense
traveling band of fSPPs disappears, and the clumps of sSPPs move randomly
in the system. The orientation distributions $P(\theta)$ corresponding to
phases (a) and (b) are shown in Fig.\ref{confi_150}(c)-I and II respectively.
In Fig.\ref{confi_150}(c)-I,
$P(\theta)$ of both the SPPs are peaked at a particular $\theta$. Thus,
the sSPPs follow the direction of motion of the dense band of the
fSPPs. As a band forms, the majority of the fSPPs travel in a particular
direction. These highly oriented fSPPs interact with the
sSPPs and influence them to orient in the same
direction, whereas, in Fig.\ref{confi_150}(c)-II, the distributions
$P(\theta)$ for both the SPPs are flat. Hence, as the band disappears,
the fSPPs get oriented randomly, making the orientation of sSPPs random.

The steady-state dynamics at $\eta=\eta_c$ of $\phi_f$ and $\phi_s$ are
given in Fig.\ref{confi_150}(d) where the values of both $\phi_f$
(blue) and $\phi_s$ (maroon) oscillate between two phases in a
synchronized manner. The higher value of $\phi_f$ corresponds to the
presence of the dense band (an ordered phase), and the lower value
of $\phi_f$ corresponds to the disappearance of the band (a
random or disordered phase). On the other hand, for the sSPPs, the
higher value of $\phi_s$ corresponds to the ordered phase of
directed clumps (DC) without forming any traveling band, and the lower
value of $\phi_s$ corresponds to a disordered phase of random clumps (RC).

\begin{figure}[t]
\centerline{
  \hfill\psfig{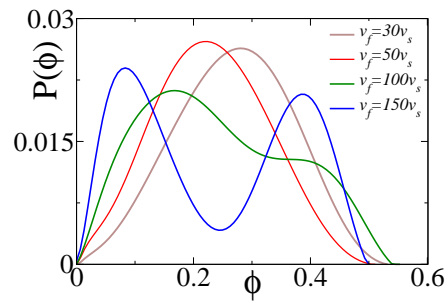}
\hfill }
\caption{\label{pd_all} Plot of $P(\phi)$ at the respective $\eta_c$s for
  different values of $v_f$ as $30v_s$, $50v_s$, $100v_s$ and $150v_s$ with
  fixed $v_s=0.01$. The system size is $L=256$.}
\end{figure}

\subsection{Discussion}
In this binary model, an orientational order-disorder phase transition
is found to occur at the respective $\eta_c$ for different velocity cases
and the transition nature is also characterized.
For both the $v_f=30v_s$ and $v_f=50v_s$ cases, a continuous transition
occurs in the system from the DS+DC phase to the RS+RC phase.
However, this is quite surprising for the fSPPs, with the velocity
$v_f=50v_s=0.5$. For the velocity $v_0=0.5$ and density $\rho_0=0.25$,
it is known that the crossover system size $L^*(\rho_0,v_0)\approx
150$ \cite{chate2008} to observe discontinuous transition in the
VM. However, in the BM, no density bands of fSPPs with $v_f=50v_s$ occur
when mixed with the sSPPs $(v_s=0.01)$ even at a system
size $L=256$ much higher than the crossover size $L^*$. It seems that
sSPPs have a significant influence on the non-formation of
traveling bands by the fSPPs as well as on the nature of
the phase transition. It is interesting to observe that both types of
SPPs maintains the same nature of transition in the BM.
Though the nature of transition with $v_f=30v_s$
and $v_f=50v_s$ remains the same as the VM at low velocities, the
values of the critical exponents, however, are different (shown in
Table.\ref{t1}). Not only are the exponents different from those in
VM, but they
are also velocity-dependent. Hence, the binary model is non-universal.

The $v_f=150v_s$ case is very different from the previous two scenarios.
In this case, the discontinuous transition was observed for both fSPPs
and sSPPs because of phase synchronization. Such a phase synchronization
occurs because of the local alignment (Eq.\ref{d1}).
Though the sSPPs show a discontinuous transition, they do not
form any traveling density band in the system. Thus, the two
different species are not able to undergo two different types of
transitions, continuous or discontinuous, simultaneously in the BM
because of the local inter-particle interaction.

A crossover from continuous to discontinuous transition is
expected at an intermediate $v_f$. This could be identified by
plotting the order parameter distribution $P(\phi)$. The distribution
will be unimodal for continuous transition and bimodal for discontinuous
transition. In Fig.\ref{pd_all}, $P(\phi)$ is plotted for different $v_f$s.
It starts deviating from the unimodal distribution near $v_f=100v_s$.
It is important to note that this crossover value of $v_f$ is determined
for the fixed system of size $L=256$ and density $\rho=0.5$.

\section{Effect of interaction radius $R$}
\begin{figure}[t]
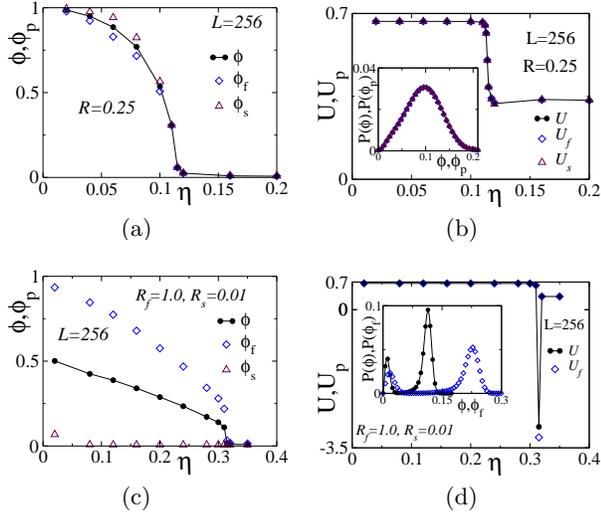

  \centerline{
    \hfill\psfig{file=fig13a.eps,width=0.21\textwidth}
       \hfill\psfig{file=fig13b.eps,width=0.21\textwidth}
  \hfill}
  \centerline{\hfill (a) \hfill\hfill (b) \hfill}
       \vspace{1.0 em}
    \centerline{
    \hfill\psfig{file=fig13c.eps,width=0.21\textwidth}
       \hfill\psfig{file=fig13d.eps,width=0.21\textwidth}
  \hfill }
\centerline{\hfill (c) \hfill\hfill (d) \hfill}
\caption{\label{R_150_256} $v_{f}=150 v_s, v_s=0.01$: Plot
  of (a) $\phi$ and $\phi_p$, (b) $U$ and $U_p$ versus $\eta$ for
  $R=0.25$. In the inset of (b), $P(\phi)$ and $P(\phi_p)$ at $\eta_c$ are
  plotted. Plot of (c) $\phi$ and $\phi_p$, (d) $U$ and $U_f$
  versus $\eta$ for $R_f=1.0$ and $R_s=0.01$. In the inset of (d), $P(\phi)$
  and $P(\phi_f)$ at $\eta_c$ are plotted. System size is $L=256$. }
\end{figure}

The value of $R$ essentially determines the number of neighbors ($n$)
to interact for an SPP at every move. So $n$ decreases as $R$ decreases.
So far, the interaction radius $R$ for both the SPPs was
set to $1$. It is important to know the effect of interaction radius
$R$ on the system with high velocity ratio $v_f=150v_s$ among two
SPPs. First, we investigate reducing the interaction radius to $R=0.25$
for both the SPPs. The order parameters ($\phi$ and $\phi_p$) and the
Binder cumulants ($U$ and $U_p$) are plotted against $\eta$ in
Fig.\ref{R_150_256}(a) and Fig.\ref{R_150_256}(b) respectively. It can
be seen that all the order parameters continuously go to zero at a
critical noise $\eta_c\approx 0.11$, which is much less than the
$\eta_c=0.33$ with $R=1$. Such a behavior is also observed in the
monodispersed case \cite{cambui2016critical}. Moreover, $U$, $U_f$,
and $U_s$ are all found to be positive. In the
inset of Fig.\ref{R_150_256}(b), $P(\phi)$ and $P(\phi_p)$ are plotted
at $\eta_c$ and all the three distributions are unimodal. Thus the
discontinuous transition for $R=1$ becomes continuous for $R=0.25$
for both the SPPs with $v_f=150v_s$, $v_s=0.01$.
Because of the reduced interaction region, very few SPPs participate
in providing the average orientation to a particular SPP, so the
coupling among the SPPs becomes weak. It has two effects. First,
because of the weak coupling, the transition
occurs at a lower value of $\eta$ ($\approx 0.11$). Second, the
fSPPs are also not able to bind the other fSPPs to form a dense band
at the transition point.

As a second case, we study the same system with different interaction
radii for two types of SPPs. Two different radii are taken as
$R_f=1.0$ and $R_s=0.01$ for the fSPPs and sSPPs
respectively. The order parameter $\phi$ and $\phi_p$ are plotted
against $\eta$ in Fig.\ref{R_150_256}(c). Note that $\phi_{s}\approx 0$
for the whole range of $\eta$, because the interaction radius is so small
that the sSPPs have almost no other particles to interact with, for this $\rho$. On the
other hand the fSPPs having $R_f=1$, undergo a discontinuous transition
at $\eta_c\approx 0.32$, slightly less than the case
with $R_f=R_s=1$ and $v_f=150v_s$, $v_s=0.01$. In the present scenario, the
sSPPs are non-interacting particles, and
the fSPPs are the only active and interacting particles in the system.
The system behavior is governed by the fSPPs only. The discontinuous
transition is then expected at $\eta_c\approx 0.32$. The $\phi$ of the
whole system is just the simple average of $\phi_s$ and $\phi_f$,
where $\phi_s$ is essentially zero. The Binder cumulants and the order
parameter distributions show the characteristics of discontinuous transition
as expected (shown in Fig.\ref{R_150_256}(d)).

Density can also be a parameter for studying this system. The results
presented above are for a fixed density $\rho=0.5$.
However, at higher densities, the traveling bands may appear at lower
velocities. Thus, one expects discontinuous transitions to occur for
lower velocity regimes at higher densities. Such systems are
computationally expensive as they involve a large number of interacting
particles.

\section{Conclusion}
A binary mixture of SPPs displays a variety of collective patterns,
such as directed lanes, clusters, clumps, micro-cluster, micro-clumps,
and others at different values of angular noise. The lane patterns of
sSPPs at a very low angular noise are supposed to be the outcome of
shear between the two species of different velocities. With a small
increase in noise, the lanes formed by the slow-moving species break
down into
clumps resulting in a drastic structural change in the system. Due to
a large difference in velocities between the two species, finite phase
segregation occurs in the system at low noise. In the
steady-state, the two species of the system are found to be
orientationally phase synchronized irrespective of their
velocities. The two-point correlation function extends throughout the
system below a critical noise. At the critical noise, the system
undergoes a velocity-dependent orientational order-disorder
transition. The nature of the transition is found to be continuous for
$v_f<100v_s$ with $v_s=0.01$ and discontinuous above this limit.
However, the continuous transitions are
non-universal as their critical exponents depend on the velocities.
The model results depend on the density of species and
the radius of interaction among the species.

\acknowledgments{
We gratefully acknowledge the computational facilities ``Newton HPC''
and ``Param-Ishan'' provided by the Department of Physics, Indian
Institute of Technology Guwahati, Assam, India.}

\bibliography{references.bib}

\end{document}